\documentclass[%
reprint,
superscriptaddress,
showpacs,
amsmath,amssymb,
prb,
]{revtex4-1}

\usepackage{graphicx}% Include figure files
\usepackage{dcolumn}% Align table columns on decimal point
\usepackage{bm}% bold math
\usepackage{color}
\usepackage[draft,inline,marginclue]{fixme}
\begin{document}
\preprint{APS/123-QED}
\title{First-principles Green's-function method for surface calculations: a pseudopotential localized basis set approach}
% Force %Force line breaks with \\
%\title{Green's function \textit{ab-initio} method for surface calculations}% Force line breaks with \\
\author{S\o{}ren Smidstrup}
\affiliation{Synopsys Inc., Fruebjergvej 3, DK-2100 Copenhagen, Denmark}
\affiliation{Faculty of Physical Sciences, University of Iceland VR-III, 107 Reykjav\'{\i}k, Iceland}
\author{Daniele Stradi}
\email{stradi@synopsys.com}
\affiliation{Synopsys Inc., Fruebjergvej 3, DK-2100 Copenhagen, Denmark}
\author{Jess Wellendorff}
\affiliation{Synopsys Inc., Fruebjergvej 3, DK-2100 Copenhagen, Denmark}
\author{Petr A. Khomyakov}
\affiliation{Synopsys Inc., Fruebjergvej 3, DK-2100 Copenhagen, Denmark}
\author{Ulrik G. \surname{Vej-Hansen}}
\affiliation{Synopsys Inc., Fruebjergvej 3, DK-2100 Copenhagen, Denmark}
\author{Maeng-Eun \surname{Lee}}
\affiliation{Synopsys Inc., Fruebjergvej 3, DK-2100 Copenhagen, Denmark}
\author{Tushar Ghosh}
\affiliation{Department of Applied Physics, Aalto University, Espoo, Finland}
\author{Elvar J\'onsson}
\affiliation{Faculty of Physical Sciences, University of Iceland VR-III, 107 Reykjav\'{\i}k, Iceland}
\affiliation{Department of Applied Physics, Aalto University, Espoo, Finland}
\author{Hannes J\'onsson}
\affiliation{Faculty of Physical Sciences, University of Iceland VR-III, 107 Reykjav\'{\i}k, Iceland}
\affiliation{Department of Applied Physics, Aalto University, Espoo, Finland}
\author{Kurt Stokbro}
\affiliation{Synopsys Inc., Fruebjergvej 3, DK-2100 Copenhagen, Denmark}
\date{\today}
\begin{abstract}
We present an efficient implementation of a surface Green's-function method for atomistic modeling of surfaces within the framework of density functional theory using a pseudopotential localized basis set approach.
In this method, the system is described as a truly semi-infinite solid with a surface region coupled to an electron reservoir,
thereby overcoming several fundamental drawbacks of the traditional slab approach.
The versatility of the method is demonstrated with several applications to
surface physics and chemistry problems that are inherently difficult to
address properly with the slab method, including metal work function calculations, band alignment in thin-film semiconductor heterostructures, surface states in metals and topological insulators, and surfaces in external
electrical fields. Results obtained with the surface Green's-function method are compared to experimental measurements and slab calculations to demonstrate the accuracy of the approach.
% HJ add
%The computational effort can be roughly an order of magnitude smaller than for slab calculations because the surface region
%in the atomic scale model needs to be only about half as large.
%
%This first-principles method describes the surface as a truly semi-infinite system in which the surface region is coupled to an electron reservoir, %and thereby overcomes several fundamental drawbacks of the traditional slab approach to surface calculations. We show the versatility of the surface %Green's-function method for a range of surface physics and chemistry problems that are inherently difficult to properly address with the slab method, %including the calculation of metal work functions, band alignment in thin-film semiconductor heterostructures, surface states in metals and %topological insulators, and properties of adsorbates interacting with surfaces in external electric fields. The results obtained with the surface %Green's-function method are compared to experiment and slab calculations, demonstrating the efficiency and accuracy of our Green's-function%% %implementation.
\end{abstract}
\pacs{71.15.-m, 31.15.E-, 73.20.-r, 68.43.-h, 68.47.Fg}
%Methods for electronic structure calculations, 71.15.-m
%density-functional theory, 31.15.E-
%Electron density of states surfaces and interfaces, 73.20.-r
%Adsorption at solid surfaces, 68.43.-h
%Semiconductors solid surfaces of, 68.47.Fg
\maketitle

\section{\label{sec:Introduction}Introduction}
Atomic-scale modeling has established itself as a workhorse tool in computational materials science.
First-principles methods are routinely applied to study the physical and chemical properties of materials and material structures, including surface structures.\cite{Fulde1995,Ceder1998,Zhang2001,Norskov2006,Wood2008,Jain2013}
The slab approach to surface calculations, which models a surface structure with just a few atomic layers, has become the \textit{de facto} standard for first-principles atomistic simulations of surfaces.
This is despite the fact that a physical surface is a semi-infinite system, interfaced to the \textit{vacuum}, unless the surface of an unsupported ultra-thin film or membrane is considered.

A slab is by construction finite in the direction perpendicular to the surface plane, and it therefore has two surfaces, which are not always equivalent. As a consequence, the electronic structure of the surfaces of the slab is altered by quantum confinement along this out-of-plane direction. It means that the accuracy of the slab approach to modeling a semi-infinite surface may critically depend on the slab thickness.\cite{lekka2003tight,martsinovich2010electronic} This leads to a number of fundamental limitations on the applicability of the slab model for surface calculations. For example, converging surface properties such as work functions and surface energies with respect to the slab thickness is notoriously difficult,\cite{Singh2009,Fall1999} and using thin slabs can result in an inaccurate electronic structure for both metal\cite{Stradi2013} and semiconductor surfaces.\cite{AliShah2012,Sagisaka2017}
This drawback is well known with the cluster approach to modeling periodic systems, where the property of interest often exhibits a slow and sometimes cumbersome convergence behavior with respect to the cluster size.\cite{teVelde1993,Tracey2013}

Different alternative methods based on the surface Green's-function (SGF) formalism have been
proposed to overcome the drawbacks of the slab approach to surface modeling.\cite{inglesfield1988surface,maclaren1989layer,skriver1991self,kudrnovsky1992self,szunyogh1994self,ishida2001surface,papior2017improvements,Wissing2013}
In the SGF method, the semi-infinite system is divided into a finite surface region and a semi-infinite bulk region, as shown in Fig.~\ref{fig:surface_configuration}.
The bulk region acts as an electron reservoir, and the surface region is coupled to this bulk region through the self-energy as discussed in Refs.~\onlinecite{Brandbyge2002,Ozaki2010,Datta1997}. The electronic structure of the entire surface system is calculated in a self-consistent manner, accounting for charge transfer between the bulk and surface regions, as well as for charge redistribution in the surface region.
It means that the surface region becomes an open system interacting with the infinite reservoir of electrons that provides a physically correct description of a semi-infinite surface structure.

In spite of their advantages, the SGF-based methods have not found broad application in computational surface science, where the slab model continues to be the method of choice.
This might be partly because one of the most popular implementations of density functional theory\cite{Hohenberg1964,Kohn1965} (DFT) is based on the pseudopotential plane-wave basis set approach,\cite{Payne1992}
which allows one to accurately converge the DFT calculations of material properties with respect to the basis set functions in a simple, systematic manner.\cite{hammer2000theoretical,kresse1996efficient} This approach is also computationally demanding for calculating large surface structures. In the  linear combination of atomic orbitals (LCAO) approach, the Kohn-Sham (KS) single-particle Hamiltonian is represented in a tight-binding-like matrix form, which can be naturally adopted within the framework of the SGF formalism. The DFT calculations done with LCAO basis sets usually have a lower computational cost compared to that of the DFT plane-wave calculations since a relatively small number of localized basis functions is employed in practical calculations. That has its downside, as the use of too few basis functions may alter the computational accuracy.

Several implementations of the surface Green's-function formalism have been recently reported for both localized\cite{cerda2012,papior2017improvements} and plane-wave basis set methods.\cite{ishida2014,olsen2016}
The latter takes advantage of a real-space representation for the Bloch states within the framework of the embedding method\cite{Inglesfield2015} or the maximally-localized Wannier-function approach.\cite{marzari2012}
The computational issues discussed in the previous paragraph still hold true for the plane-wave and LCAO-based SGF implementations, and need to be properly addressed to allow for both efficient and reliable SGF-based surface calculations.

In this paper, we present an efficient, accurate, self-consistent SGF method for first-principles calculations of the total energy and electronic structure of surfaces that has been implemented in the Atomistix ToolKit (ATK) simulation tool within the framework of the DFT pseudopotential LCAO basis set approach.\cite{Stradi2016,ATK} The present implementation of the SGF method takes an advantage of the highly-optimized Green's-function methodology that has already been implemented in the ATK code for two-probe device simulations.\cite{Brandbyge2002,Soler2002} We develop new optimized LCAO basis sets (see Appendix) used in combination with recently-developed SG15 optimized norm-conserving Vanderbilt pseudopotentials.\cite{Schlipf2015} This allows for highly-accurate LCAO calculations of material structure properties, with an accuracy similar to that of plane-wave based methods, and the computational efficiency of LCAO-based methods. This is of particular importance for an accurate description of the surface structures studied in our work.

We apply the ATK-SGF method to several surface problems that are inherently difficult to properly address with the traditional slab approach, including the calculation of metal work functions, band alignment in thin-film semiconductor heterostructures, surface states in metals and topological insulators, and the properties of adsorbates interacting with surfaces in external electric fields.
For these studies, the ATK-SGF implementation has been combined with several methodological developments:
(i) a real-space multigrid approach for imposing non-periodic boundary conditions, e.g., for work function calculations or surface calculations with external electric field,
(ii) an implementation of doping methods, e.g., for modeling doped semiconductor substrates,\citep{Stradi2016}
(iii) a pseudopotential projector-shift method for resolving the problems of DFT in describing correctly the band gap of semiconductors (see Appendix),
(iv) an implementation of spin-orbit coupling, which is an important effect in topological insulators,\citep{Chang2015} and
(v) self-consistent total energy calculations directly within the SGF method, e.g.,
for studying adsorbates on the metal surfaces.

The paper is organized as follows. Section~\ref{sec:methodology} describes the methodology and basic computational settings adopted in this work, as well as implementation details of the SGF method and its computational efficiency. Section~\ref{sec:WFResults} shows how to calculate work functions of metal surfaces that are well-converged with respect to the system size, using the SGF method. In Sec.~\ref{sec: Band alignment}, the SGF method is applied for understanding of the band alignment in a semiconductor heterostructure such as a Si film on intrinsic and doped Ge(001) substrates. Section~\ref{sec:SS} shows how the SGF method can be used to calculate pure surface states in metals and topological insulators. Section~\ref{sec:surface_chemistry} describes an application of the SGF method for surface chemistry problems such as the adsorption of iodine atoms on the Pt(111) surface in the presence of an external electric field. The main conclusions are summarized in Sec.~\ref{sec:Conclusions}.

\section{\label{sec:methodology}Methodology}
\subsection{Electronic structure method}
Our implementation of the surface Green's-function method is done within the framework of
density functional theory\cite{Hohenberg1964,Kohn1965,Kohn1996,Parr1994}
using the norm-conserving pseudopotential LCAO basis set approach.\cite{Hamann1979,Soler2002}
The corresponding Kohn-Sham (KS) Hamiltonian can be written as
\begin{equation}
\hat{H}^\text{KS} =  -\frac{\hbar^2 }{2m} \nabla^2 + V_\text{loc}
                     + V_\text{nl} + V_\text{H} + V_\text{xc} ,
\label{eq:HIJ2}
\end{equation}
where the first term corresponds to the electron kinetic energy, $V_\text{loc}$ and $V_\text{nl}$ are the local and nonlocal parts of the pseudopotential, respectively, and the Hartree ($V_\text{H}$) and exchange-correlation ($V_\text{xc}$) potentials are given by the last two terms.

Using an LCAO basis allows representing the KS Hamiltonian in a matrix form with the following matrix elements\cite{Soler2002,Brandbyge2002}
\begin{equation}
\mathrm{H}^\text{KS}_{ij} = \langle \phi_i(r) | \hat{H}^\text{KS} | \phi_j(r) \rangle ,
\label{eq:HIJ1}
\end{equation}
where $\phi_{i}(r)$ and $\phi_{j}(r)$ are localized finite-range numerical orbitals.\cite{Junquera2001,Ozaki2003} To evaluate the Hamiltonian matrix elements in Eq.~\eqref{eq:HIJ1}, we follow the SIESTA method,\cite{Soler2002}
where the $V_\mathrm{H}$ and $V_\mathrm{xc}$ terms are calculated on a real-space grid.
\subsection{Pseudopotentials and basis sets}
Using a pseudopotential LCAO approach requires a careful choice of the pseudopotential and LCAO basis set to do computationally-efficient DFT calculations without compromising the accuracy of the obtained numerical results. We have developed three types of SG15 pseudopotential-based basis sets corresponding to Ultra, High and Medium accuracy for all elements in the periodic table up to $Z=83$. The SG15-Ultra basis sets provide the accuracy of DFT-LCAO calculations comparable to that of the state-of-the-art all-electron calculations, whereas the SG15-Medium basis set type allows for computationally-cheap calculations with an error that is of the same order as that due to the use of approximate DFT functionals within the framework of local density (LDA) or generalized gradient approximations (GGA). Adopting the Medium basis set, we typically gain an order of magnitude in the computational efficiency compared to the Ultra basis set. In the Appendix, we present the methodology to generate these basis sets, and benchmark the corresponding DFT-LCAO calculations against reference all-electron and pseudopotential plane-wave DFT calculations to evaluate the pseudopotential and basis set accuracy.

A reliable study of semiconductor physics problems usually requires an accurate description of the band gap. Unfortunately, the DFT approach based on local and semi-local DFT density functionals fails to accurately calculate the band gap of semiconductor materials.\cite{MoriSanchez2008} To overcome this problem we have introduced a set of adjustable parameters for the pseudopotentials somewhat similar to the empirical pseudopotentials proposed by Zunger and
co-workers.\cite{Wang1995} This approach allows for a good description of both structural and electronic properties of semiconductors. This method has been used for studying a Si thin-film on the Ge(001) substrate in Sec.~\ref{sec: Band alignment} (more details on the generation of the parameters can be found in the Appendix).
\subsection{Green's-function method}
\begin{figure*}
\includegraphics[scale=0.125]{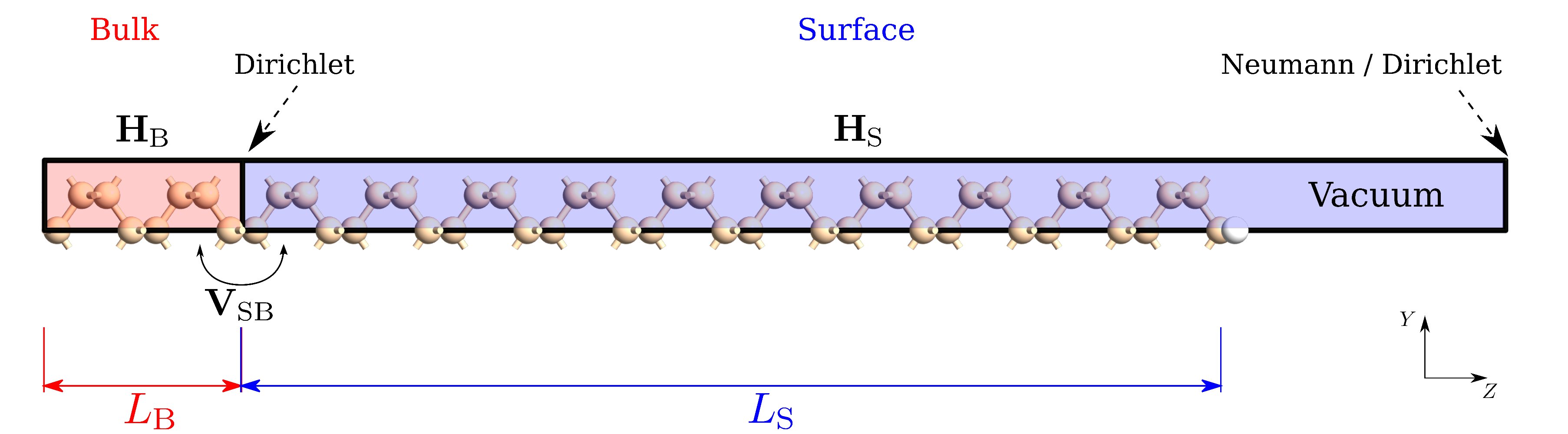}
\caption{Illustration of a typical semi-infinite surface configuration.
The Hamiltonians of the surface region ($\mathbf{H}_\mathrm{S}$) and the principle layer of the
semi-infinite bulk region ($\mathbf{H}_\mathrm{B}$), are coupled through the coupling terms $\mathbf{V}_\mathrm{SB}$. A Dirichlet
boundary condition is used at the boundary between the bulk and
surface regions. In the vacuum region, a Neumann (Dirichlet) boundary
condition is used for calculations without (with) an electric field. $L_\mathrm{S}$
($L_\mathrm{B}$) is the length of the surface region (the
bulk region principle layer) along the out-of-surface-plane ({\it Z}) direction.}
\label{fig:surface_configuration}
\end{figure*}
Using a finite-range LCAO basis set allows for partitioning the Hamiltonian of the semi-infinite surface into three distinct matrix blocks that correspond to the Hamiltonian of the surface region ($\mathbf{H}_\mathrm{S}$), a single atomic layer (``principal layer'') of the semi-infinite bulk region ($\mathbf{H}_\mathrm{B}$) and the coupling matrices ($\mathbf{V}_\mathrm{BB}$ and $\mathbf{V}_\mathrm{SB}$), as illustrated in Fig. \ref{fig:surface_configuration}.\cite{Brandbyge2002} The coupling matrices, $\mathbf{V}_\mathrm{SB}$ and $\mathbf{V}_\mathrm{BB}$, account for interaction between the surface and bulk region atomic layers, and between the principal layers of the semi-infinite bulk region, respectively. In the ATK implementation, the coupling matrix, $\mathbf{V}_\mathrm{SB}$, is expressed in terms of the $\mathbf{V}_\mathrm{BB}$ matrix as described in Ref.~\onlinecite{Brandbyge2002}, assuming that a sufficiently thick layer of the material comprising the semi-infinite bulk region is added to the surface region. The infinite Hamiltonian matrix of the entire system can then be written as
\begin{equation}
\mathbf{H}^\mathrm{KS} =
 \begin{pmatrix}
 \ddots & \vdots                             & \vdots                           & \vdots                           & \vdots                 \\
 \ldots & \mathbf{V}_\mathrm{BB}^{\dagger}   & \mathbf{H}_\mathrm{B}            & \mathbf{V}_\mathrm{BB}           & 0                      \\
 \ldots & 0                                  & \mathbf{V}_\mathrm{BB}^{\dagger} & \mathbf{H}_\mathrm{B}            & \mathbf{V}_\mathrm{SB} \\
 \ldots & 0                                  & 0                                & \mathbf{V}_\mathrm{SB}^{\dagger} & \mathbf{H}_\mathrm{S}
 \end{pmatrix}.
\end{equation}
Using Green's-function formalism,\cite{Haug2008} the density matrix of the surface region,
$\mathbf{D}_\mathrm{S}$, can be expressed as
\begin{equation}
\mathbf{D}_\mathrm{S} = -\frac{1}{\pi}\int_{-\infty}^{\mu_\mathrm{B}}
                \mathrm{Im}[\mathbf{G}_\mathrm{S}(\epsilon)]d\epsilon ,
\label{eq:negf_1}
\end{equation}
where $\mu_\mathrm{B}$ is the bulk chemical potential, and $\mathbf{G}_\mathrm{S}$ is the finite Green's-function matrix of the surface region
\begin{equation}
\mathbf{G}_\mathrm{S}(\epsilon) = [(\epsilon + i\delta)\mathbf{S}_\mathrm{S}-\mathbf{H}_\mathrm{S}-\mathbf{\Sigma}(\epsilon)]^{-1},
\label{eq:negf_2}
\end{equation}
where $\mathbf{S}_\mathrm{S}$ and $\mathbf{H}_\mathrm{S}$ are the overlap and Hamiltonian matrices
associated with the basis set functions centered inside the surface region, respectively;
$\mathbf{\Sigma}$ is the self-energy matrix describing the coupling of the
surface to the semi-infinite bulk region, i.e., accounting for open boundary conditions imposed on the surface region. In most cases, the initial guess for the Hamiltonian $\mathbf{H}_\mathrm{S}$ can be constructed from a superposition of atomic densities. Obtaining the initial guess for $\mathbf{H}_\mathrm{S}$ from a conventional calculation of a slab corresponding to the surface region is also possible for systems exhibiting difficult convergence behavior, which is the case of the calculation including non-collinear spin-orbit coupling carried out in this work for the Bi$_2$Se$_3$(111) surface, presented in Section \ref{sec:SS}B.

Given the density matrix, the electron density, $n(r)$, is constructed as
\begin{equation}
n(r) = \sum_{ij} \left[ D_{\mathrm{S}} \right]_{ij} \phi_i(r) \phi_j(r) + n_\mathrm{sp},
\label{eq:realspacedensity}
\end{equation}
where $n_\mathrm{sp}$ is the ``spill-in" corrective term related to density matrix components in the bulk region and the bulk--surface boundary.\citep{Ozaki2010,Stradi2016} Including this term is crucial to describe correctly the charge density at the boundary between the surface and bulk regions, by accounting explicitly for the density in the surface region due to those basis functions in the bulk region, which tails penetrate into the surface region. For a more extensive description, we refer the reader to Ref. \citenum{Stradi2016}. The electron density, the Hartree and exchange-correlation potentials and the surface Green's function can then be obtained by solving the Kohn-Sham and Poisson equations together with Eqs.~\eqref{eq:HIJ2}--\eqref{eq:realspacedensity} in a self-consistent manner, using a procedure equivalent to that described in Ref. \citenum{Brandbyge2002}, but for a system formed by a central region coupled to a single electron reservoir. Depending on the actual physical problem of study, the Poisson equation can be solved with the Dirichlet, Neumann or mixed boundary conditions as shown in Fig.~\ref{fig:surface_configuration}.
\subsection{Implementation details}
\begin{figure}
\includegraphics[width=\linewidth]{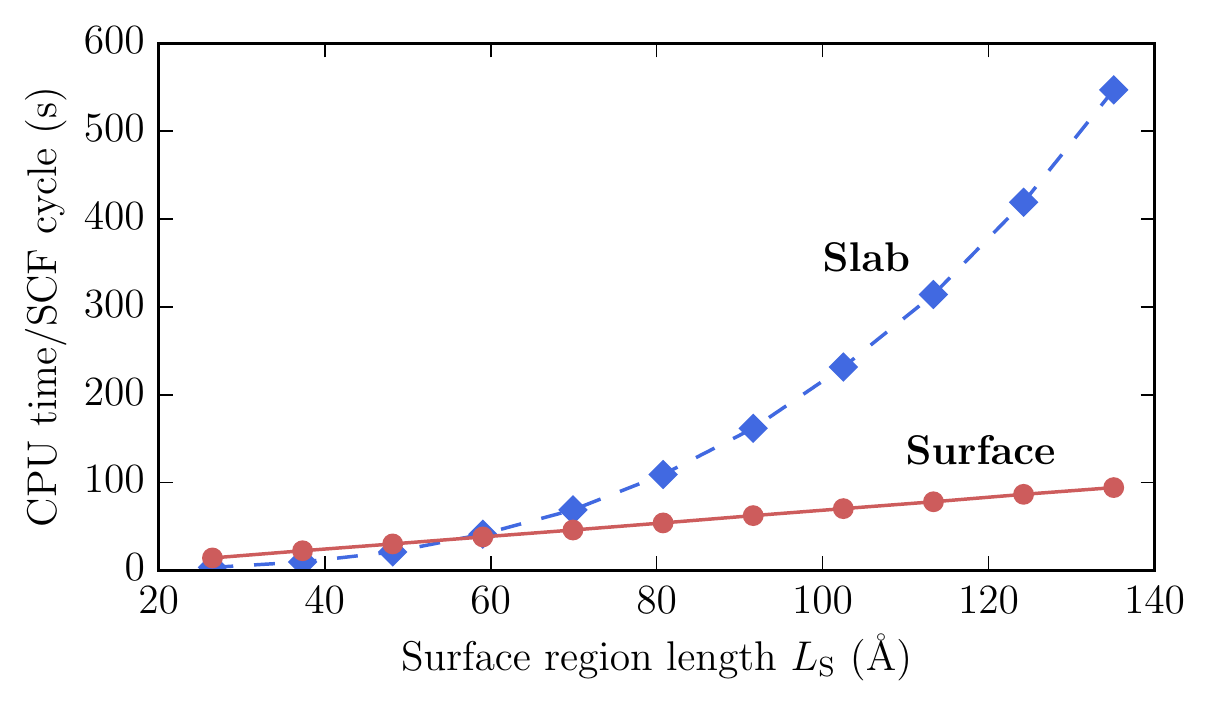}
\caption{CPU time per self-consistent (SCF) cycle as function of the length of the
surface region $L_\mathrm{S}$ (see Fig.\ref{fig:surface_configuration})
for a surface configuration (red filled circles, dashed line), compared to the CPU time
per SCF cycle of a slab configuration (blue filled diamonds, solid line) with the slab thickness equivalent to $L_\mathrm{S}$. The system
considered is a $2\times 2$ unreconstructed Si(100) surface.}
\label{fig:scaling}
\end{figure}
The numerical implementation of the SGF method is an extension of the development done for simulating two-terminal devices in the ATK.\cite{Brandbyge2002,Ozaki2010} In the SGF method, a single electron reservoir is only needed to impose the open boundary condition on the surface region. That means that the integral in Eq.~(\ref{eq:negf_1}) comprises only the equilibrium part of the Green's function, which can be efficiently evaluated using complex contour integration. Subsequently, the density matrix in Eq.~(\ref{eq:negf_1}) can be written as
\begin{equation}
\mathbf{D}_\mathrm{S} = \sum_k w_k \mathbf{G}_\mathrm{S} (z_k) ,
\end{equation}
where the complex energies $z_k$ and the weights $w_k$ are determined as described elsewhere.\cite{Brandbyge2002, Ozaki2010}

To calculate the Green's-function matrix, $\mathbf{G}_\mathrm{S}$, we have to compute the self-energy matrix ($\mathbf{\Sigma}$) of the semi-infinite bulk region, which will be called the electrode in the following.
For that, we first obtain the electrode matrices, $\mathbf{H}_\mathrm{B}$ and $\mathbf{V}_\mathrm{BB}$, from a bulk calculation using periodic boundary conditions.
The self-energy matrix in Eq.~\eqref{eq:negf_2} can then be computed directly from propagating and evanescent modes,\cite{Sanvito1999,Khomyakov2005} which can be efficiently calculated with an iterative method as proposed in Ref.~\onlinecite{Sorensen2008}. Here we adopt a more efficient recursive method for the self-energy matrix calculation that does not require an explicit calculation of the electron modes in the bulk electrode.\cite{LopezSancho1985}
Using the recursion method proposed in Ref.~\citenum{LopezSancho1985},
we exploit the sparsity of the bulk Hamiltonian matrix, and find that this method gives the best balance between
stability, accuracy, and computational efficiency.

The Green's-function matrix is eventually calculated with the Sweep method
optimized for application to the surface configuration.\cite{petersen2008block}
This method allows for finding the Green's-function matrix in $O(N)$ steps,
where $N$ is the number of diagonal blocks in the block tridiagonal Hamiltonian matrix.
The Hamiltonian matrix elements are preordered
to give an optimal block tridiagonal structure.\cite{papior2017improvements}
Alternatively, the MUMPS\cite{amestoy2001fully} and PEXSI\cite{Lin2013} libraries, which allow for lower memory consumption and parallel scaling to a larger number of computing processors, can also be employed. We find, however, that their serial performance is worse than that of the Sweep method, in general.

A significant advantage of using Green's-function techniques is that the complexity of the calculation scales as $O(M^{\alpha} N)$ instead of the typical $O(M^{3} N^{3})$ scaling of DFT calculations using periodic boundary conditions, where $2<\alpha\leq 3$, and $M$ is the dimension of the matrix corresponding to each of the $N$ blocks in the block tridiagonal Hamiltonian matrix. The actual value of $\alpha$ depends on the particular implementation of matrix operations adopted for Green's-function matrix calculations. The time required for a single Green's-function SCF cycle therefore scales linearly with the number of surface atomic layers, instead of the usual cubic scaling. Figure~\ref{fig:scaling} shows a comparison of the CPU time per self-consistent cycle needed to calculate a surface configuration of length $L_\mathrm{S}$ and a slab configuration having an equivalent length. For a length $L_\mathrm{S} = 13.5\ \mathrm{nm}$, corresponding approximately to the width of the depletion layer in bulk silicon at an $n$-doping level of $n = 10^{18}\mathrm{cm}^{-3}$, one can see that a slab calculation is more computationally-expensive than a SGF calculation by a factor of 5.

The Hartree potential term $V_\mathrm{H}$ in Eq.~\eqref{eq:HIJ2}\ is obtained by solving the Poisson equation with a Dirichlet boundary condition at the electrode-surface interface and a Neumann boundary condition in the vacuum. These mixed boundary conditions are exact for a semi-infinite surface in the absence of an external electric field. External fields can be included by imposing Dirichlet boundary conditions also in the vacuum region, enabling simulations of surface structures in external electric fields. In both cases, the Poisson equation is solved
using either a multigrid solver or the two-dimensional (2D) FFT method introduced in Ref.\citenum{Ozaki2010}.

All time-demanding steps are parallelized in the ATK, including calculation of the Green's-function matrix in Eq.~\eqref{eq:negf_2},
the real-space density in Eq.~\eqref{eq:realspacedensity}, the real-space potentials in Eq.~\eqref{eq:HIJ2}, and Hamiltonian in Eq.~\eqref{eq:HIJ1}.
In particular, the SGF calculations are parallelized over $k$-points and contour integration points for the Green's-function matrix calculation.
\subsection{\label{sec:compdetails}Computational details}
In this paper, the ATK-DFT calculations have been done using the GGA-PBE
exchange-correlation functional\cite{Perdew1996} and the SG15-Medium
combination of norm-conserving pseudopotentials and LCAO basis sets, unless otherwise stated.
We have adopted a real-space grid density that is equivalent to a plane-wave kinetic energy
cutoff of 100~Ha, and the Monkhorst--Pack $k$-point grids for the Brillouin zone sampling.\cite{MonkhorstPack1976} For the bulk electrodes, three-dimensional grids have been used to sample the 3D Brillouin zone. In order to properly converge the self-energy matrices $\mathbf{\Sigma}$ entering in Eq.~\eqref{eq:negf_2}, very dense grids have been used in the direction normal to the surface plane.\cite{Brandbyge2002} For the SGF calculations, the system is periodic only along the directions parallel to the surface plane, so that 2D grids have been used in this case. The choice of the actual k-point sampling depends on the system considered, and will be reported in each of the following sections.
The broadening of the Fermi--Dirac distribution is chosen to be of 0.026~eV. The total energy and forces have been converged
at least to $\sim 10^{-4}$~eV and $0.01$~eV/\AA , respectively.

\section{\label{sec:WFResults}Work function calculations}
\begin{figure}
\includegraphics[width=\linewidth]{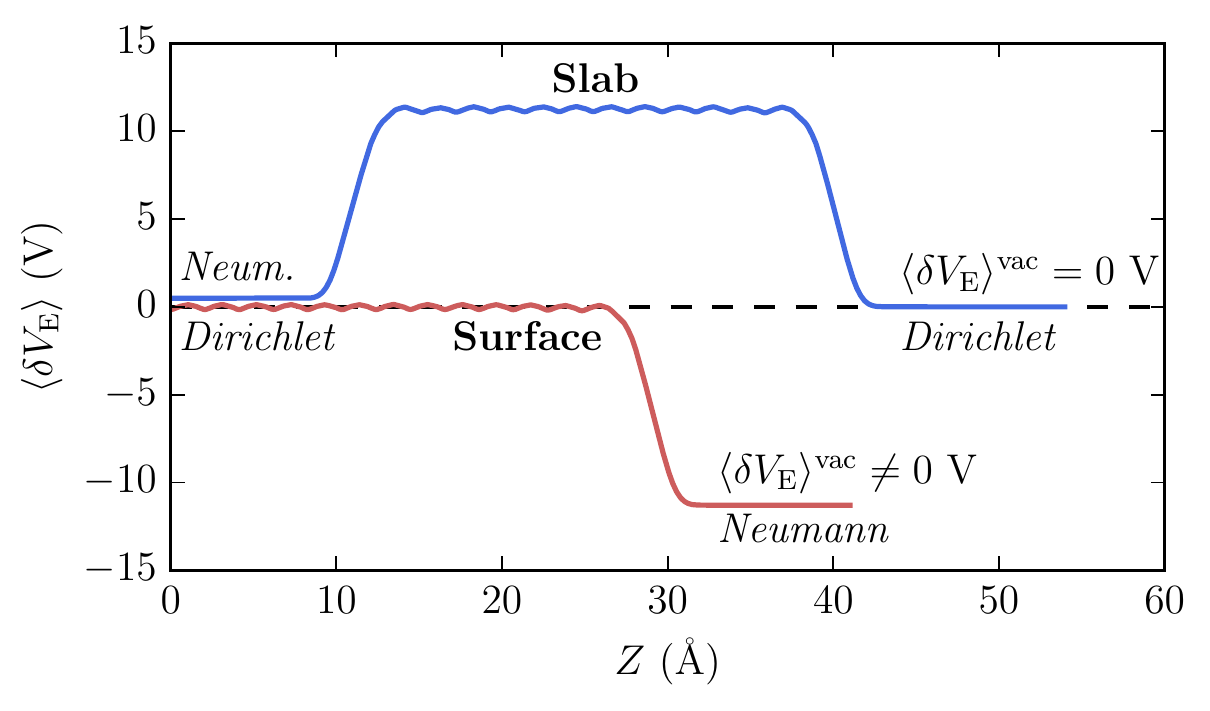}
\caption{The macroscopic in-plane averaged electrostatic difference potential, $\langle \delta V_\mathrm{E} \rangle$,\cite{EDPnote} calculated throughout the surface and slab structures for the SGF model (blue solid line) and the ATK slab model (red solid line), respectively.
For both surface models, the $\langle \delta V_\mathrm{E} \rangle$ potential is calculated for 14-monolayer Ag(001) slab or surface region, and is projected onto the $Z$-axis (normal to the surface plane). Boundary conditions of Dirichlet or Neumann type
are indicated for both SGF and slab models of the surface.}
\label{fig:WF0}
\end{figure}
\begin{figure}
\includegraphics[width=\linewidth]{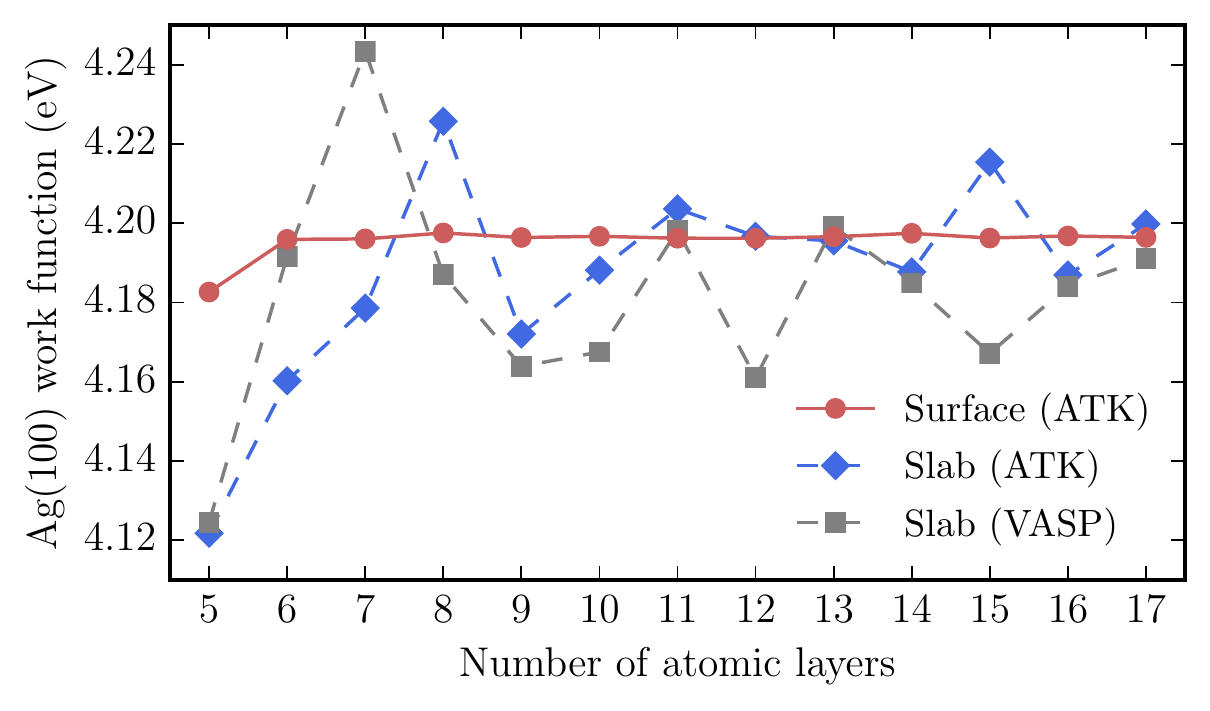}
\caption{The Ag(001) work function calculated as a function of a number of atomic monolayers in the surface region, using the ATK-SGF method (red filled circles), the ATK (blue filled diamonds) and VASP (black filled squares) slab model.}
\label{fig:WF1}
\end{figure}
\begin{figure}
\includegraphics[width=\linewidth]{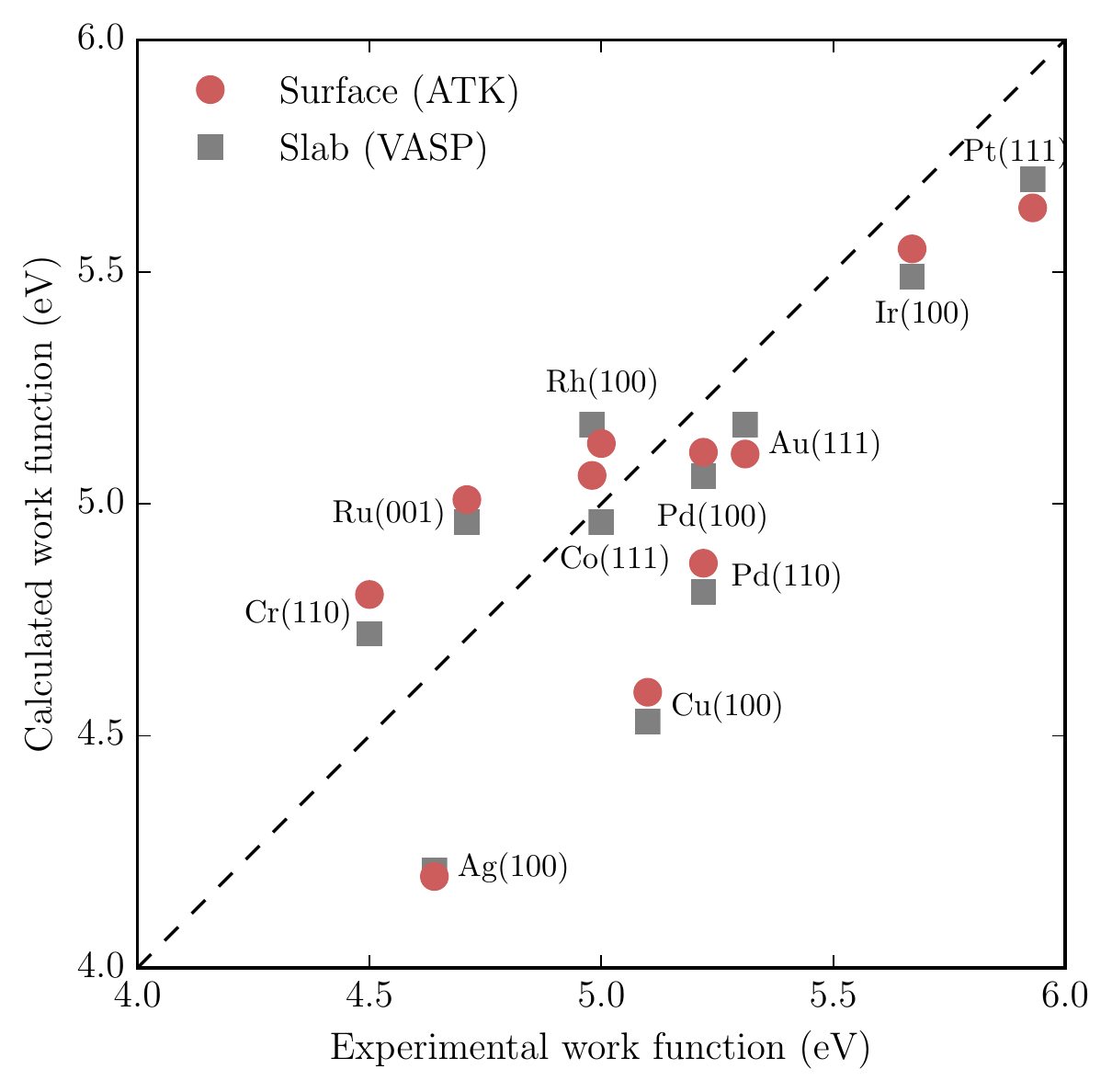}
\caption{DFT-calculated vs. experimental work functions. The work functions are calculated using the ATK-SGF method (red filled circles) and the VASP plane-wave slab method (black filled squares). The measured work functions are taken from Ref.~\onlinecite{crc2017}.}
\label{fig:WF2}
\end{figure}
The work function, $W$, is a fundamental electronic property of a surface. Knowing the work function values for metal surfaces is of
particular importance in electronics\cite{Giovannetti2008} and (photo)electrochemistry.\cite{Trasatti1971} The work function is the energy required to remove an electron from the Fermi level ($E_\mathrm{F}$) of a cleaved crystal to the vacuum level,
\begin{equation}
W = -e\phi - E_\mathrm{F} ,
\end{equation}
where $e$ is the elementary charge, $e>0$, and $\phi$ is the electrostatic potential
in the vacuum region near the surface plane.

Work function calculations based on the DFT approach most often employ a slab model
for the surface structure. This often requires using a dipole correction
to eliminate a spurious interaction between periodically repeated slab images.\citep{Neugebauer1992}
Furthermore, the computed work function may converge
slowly with respect to the number of atomic monolayers in the slab. So, accurate work
function calculations can be computationally intensive within the framework of the slab approach.
In this section, we demonstrate that employing the ATK-SGF based approach for calculating
work functions of metal surfaces resolves these issues.

\paragraph*{Methods} The fundamental difference between the slab and SGF methods for ATK-DFT surface
calculations is illustrated in Fig.~\ref{fig:WF0}, which shows the macroscopic in-plane averaged electrostatic difference potential, $\langle \delta V_\mathrm{E} \rangle$,\cite{EDPnote} throughout the
Ag(001) surface structure for the slab model and the SGF model of the surface.
Both model structures of the Ag(001) surface are effectively comprised of 14 atomic monolayers.
There exists, however, a crucial difference between the slab and SGF-modeled surface structures, as the SGF-modeled Ag(001) surface region
is matched to that of bulk Ag region as discussed in Sec.~\ref{sec:methodology}.

For work function calculations, we impose a Dirichlet (Neumann) boundary condition on the right (left) side of the slab, see Fig.~\ref{fig:WF0}. It means that the electrostatic potential is zero near the surface in the vacuum on the right side of the slab, and the slab-calculated work function ($W_\mathrm{slab}$) of the corresponding surface is given by the slab chemical potential, $E_\mathrm{F}^\mathrm{slab}$,
\begin{equation}
W_\mathrm{slab} = -E_\mathrm{F}^\mathrm{slab} .
\end{equation}
For the SGF model, a Neumann (Dirichlet) boundary condition is adopted in the vacuum (at the interface between the surface and bulk regions) as shown in Fig.~\ref{fig:WF0}. In this case, the chemical potential of the entire surface system is
that of the bulk region, and the SGF-calculated work function, $W_\mathrm{SGF}$, is then given as
\begin{equation}
W_\mathrm{SGF} = -e \langle\delta V_\mathrm{E}\rangle^\mathrm{vac}
                - E_\mathrm{F}^\mathrm{bulk} ,
\end{equation}
where $E_\mathrm{F}^\mathrm{bulk}$ is the Fermi level of the bulk region
and $\langle\delta V_\mathrm{E}\rangle^\mathrm{vac}$ is the macroscopic in-plane averaged
electrostatic difference potential near the surface in the vacuum. \citep{Baldereschi1988}

For the sake of comparison, we have calculated work functions using both the SGF and slab method. The slab model has been employed within the framework of the LCAO and plane-wave (PW) based approaches as implemented in the ATK and VASP codes, respectively.\citep{ATK,Kresse1996} We have used a $1\! \times\! 1$ surface primitive cell and
vacuum layers with a thickness of $\sim 12$~\AA . The 2D Brillouin zone (BZ) of the surface
has been sampled with a $15\! \times\! 15$ $k$-point grid, and
a $15\! \times\! 15\! \times\! 101$ $k$-point grid has been adopted for sampling 3D BZ of the bulk metal.
We have done ion relaxation for the top layers of the metal surface, converging the forces to a maximum value of 0.01 eV/\AA .
For the ATK work function calculations, three ghost atoms have been added to the surface structure near the surface to accurately
account for the electron density decaying into the vacuum.\citep{GarciaGil2009} All the other ATK computational
details are given in Sec.~\ref{sec:compdetails}. For the VASP calculations, we have
employed a kinetic energy cut-off of 400~eV and a dipole correction\citep{Neugebauer1992} in the out-of-surface-plane direction.

\paragraph*{Results} Figure~\ref{fig:WF1} shows how the Ag(001) work function, which is calculated using the slab (SGF) model, converges with respect to the number of atomic monolayers in the slab (surface region). This figure suggests that rather thick slabs
are needed to converge the work function, whereas the SGF-calculated work function is almost independent of the surface region thickness. The main reason for this fast convergence is that the SGF-calculated electronic structure of the surface region is coupled
to that of the semi-infinite bulk region, meaning that the bulk states are taken into account in an exact manner for any thickness of the surface region. In the slab approach, one would have to increase the slab thickness significantly to accurately describe the bulk states as seen in Fig.~\ref{fig:WF1}.

To demonstrate that the SGF method for work function calculations is accurate for various metal surfaces, we have computed the work functions of 11 transition metal surfaces such as the Ag(100), Au(111), Co(111), Cr(110), Cu(100), Ir(100), Pd(100), Pd(110),
Pt(111), Rh(100), and Ru(001) surface. For the work function calculations, we have built metal slabs and surface
regions with the thickness of 13 atomic monolayers, using experimental lattice parameters of bulk metals.

Figure~\ref{fig:WF2}
shows that the work functions calculated with the ATK-SGF and PW-slab approaches agree with the experimental data within a mean error of $\sim 0.26$~eV and an absolute error of $\sim 0.5$ eV.\cite{crc2017} This figure also suggests that the work function values calculated with the PW-slab approach are in a good agreement with the SGF-obtained work functions, provided sufficiently-thick (13 atomic monolayers) slabs are adopted for the slab calculations. The absolute (mean) error between the SGF- and slab-calculated work functions is in the range of $\sim 0.1$~eV ($\sim 0.07$~eV), which is smaller than the computational absolute (mean) error $\sim 0.5$~eV ($\sim 0.26$~eV) estimated by comparing the calculated work functions to measured ones in Fig.~\ref{fig:WF2}.

In conclusion, we demonstrated that using the SGF method for work function calculations is more advantageous, compared to the slab method, as the SGF-calculated work function converges much faster with respect to the thickness of the surface model structure. The ATK-LCAO results obtained in this section suggested that the ATK-LCAO basis sets (see Appendix) combined with the SG15 optimized norm-conserving pseudopotentials\cite{Schlipf2015} provide the accuracy of LCAO-based work function calculations that is similar to that of PW-based calculations.

\section{Band alignment in semiconductor heterostructures} \label{sec: Band alignment}

In this section, we address the issue of how to calculate the electronic structure of a semiconductor surface in an accurate manner, and how the band alignment between the surface and a semiconducting thin-film can then be defined and calculated from first-principles atomistic simulations. We demonstrate that the SGF approach resolves several severe limitations of the slab approach for the band structure calculations of semiconductor surfaces and interfaces.

\subsection{Ge(001) surface} \label{sec: Ge surface}

First, we study the Ge(001) surface, using the SGF approach and comparing it to the conventional slab approach. The band structure calculation of semiconductor surfaces is a challenging computational problem compared to that of metal surfaces since, among other effects, the semiconductor energy gap has a strong dependence on the slab thickness because of quantization effects. We show that the SGF approach allows one to overcome this particular drawback of the slab approach, accounting for the bulk semiconductor states in an exact manner by imposing open-boundary conditions on the semiconductor surface structure.

\paragraph*{Methods} \label{sec:MethodsBA}
For the slab calculations of the Ge(001) surface, we have built a set of Ge(001) slabs with increasing thicknesses, $L/a$ = 3, 4, 5, 6 and 7, where the lattice constant of bulk Ge optimized at the DFT level is $a = 5.725\ \mathrm{\AA}$. The two Ge(001) surfaces of each slab are passivated with hydrogen atoms to saturate the Ge dangling bonds and remove any localized surface band emerging in the band gap of Ge. A vacuum layer with a thickness of 16 \AA\, is added to separate the neighboring slab images. The Brillouin zone (BZ) has been sampled using an $8 \times 8 \times 1$ $\Gamma$-centered k-points grid.\citep{MonkhorstPack1976} The energy gap of the slab has been obtained by calculating the local density of states (LDOS) at the innermost position of the slab, using a $24\times 24 \times 1$ k-point grid for the 2D BZ, and by taking the energy difference between the highest energy occupied state and the lowest energy unoccupied state in the calculated LDOS. We have adopted the SG15-High combination of norm-conserving pseudopotential and LCAO basis set for germanium. The total energy has been converged to $\sim 10^{-5}$ eV at least.  Periodic boundary conditions are imposed in both the in-plane and out-of-plane directions.

For the SGF calculations of the Ge(001) surface, we have attached a semi-infinite bulk region to each of the Ge(001) slabs discussed in the previous paragraph, after removal of the passivating hydrogen atoms on the contacted side. We impose the Dirichlet boundary condition at the boundary located at $z=0$ \AA\ between the surface and bulk regions as shown in Fig.~\ref{fig:ge_si_hpot}a, and the Neumann boundary condition at the boundary located at the distance of 16 \AA\ above the Ge(001) surface in vacuum. All the other computational settings are adopted as for the Ge(001) slab calculations. The LDOS has been calculated at the boundary between the surface region and the bulk electrode, using a $24\times 24$ k-points grid to sample the 2D BZ of the Ge(001) surface, and by taking the energy difference between the highest energy occupied state and the lowest energy unoccupied state in the calculated LDOS.

Note that we have not performed any ion relaxation for neither slab nor SGF model of the Ge(001) surface intentionally, keeping the Ge(001) surface structure the same in both slab and SGF calculations. That allows us to separate the effect of the slab finite size from the effect of the ion relaxation on the band structure of the Ge(001) surface.

\paragraph*{Results} \label{sec:ResultsBA}

\begin{figure}
    \includegraphics[width=\linewidth]{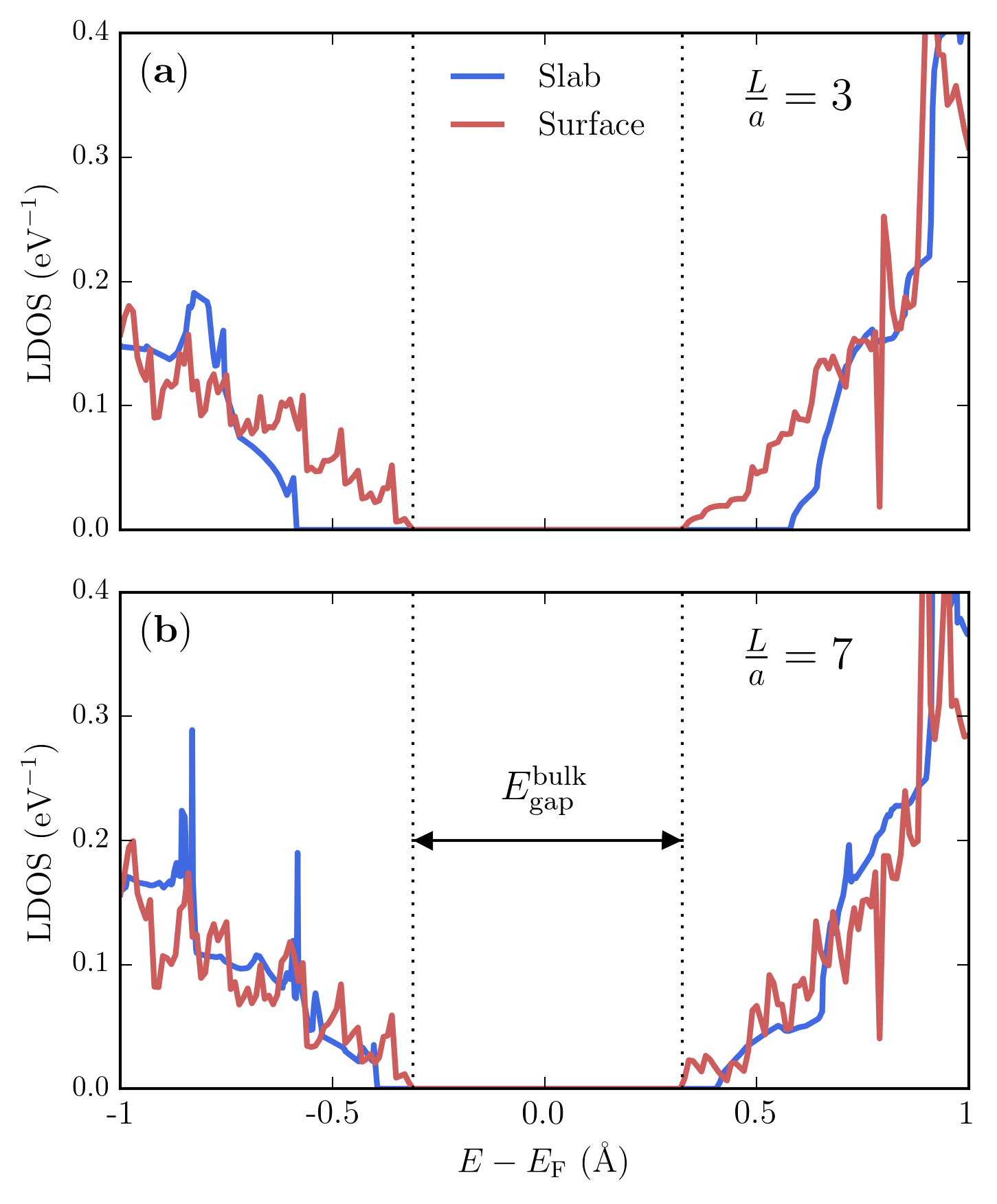}
    \caption{(a) Local density of states (LDOS) calculated for a slab configuration (blue) or a surface configuration (red) with thickness $L = 3a$, where the thickness $L$ is given in units of the lattice constant of bulk Ge, $a$. (b) Same as (a), but for $L = 7a$. The dashed vertical lines indicate the extremities of the band gap of bulk Ge.}
\label{fig:figure_BA4}
\end{figure}

\begin{figure}
    \includegraphics[width=\linewidth]{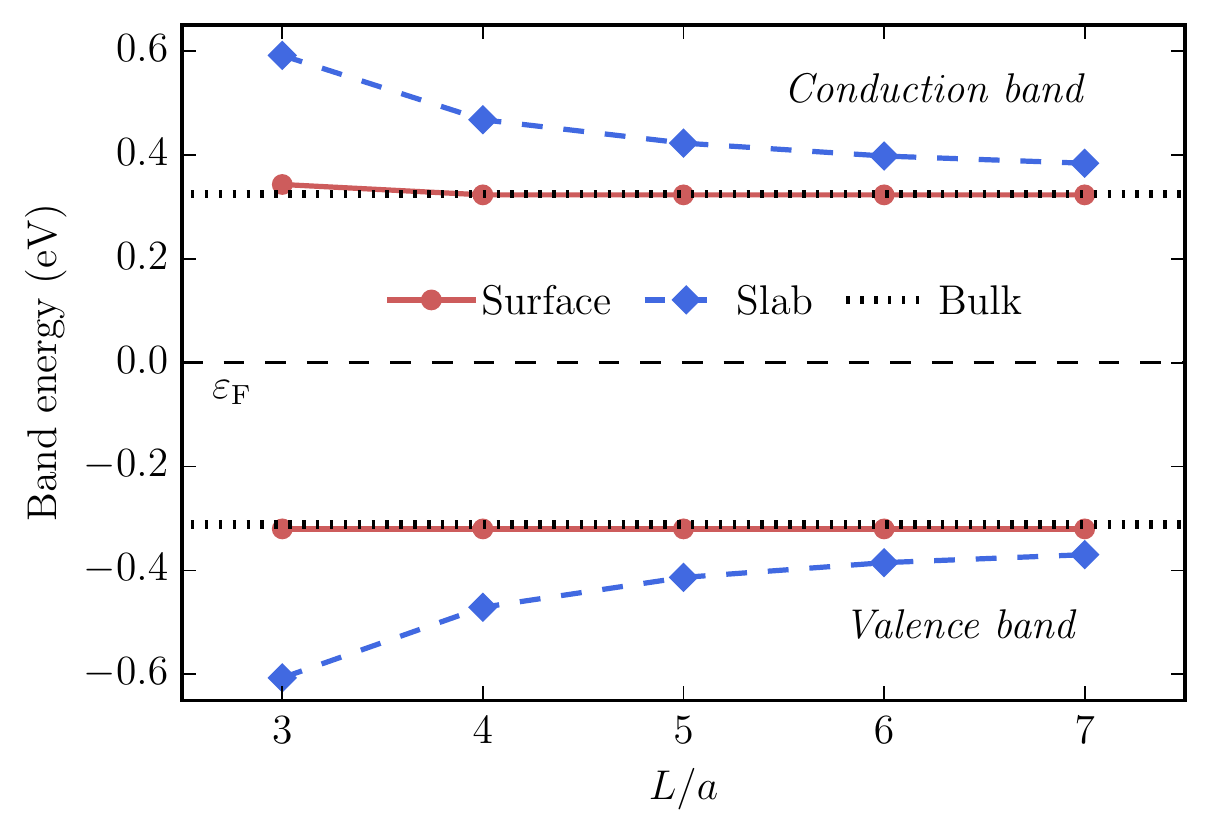}
    \caption{The conduction band minimum (green lines) and valence band maximum (blue lines) calculated for bulk Ge (solid lines), the Ge(001) surface (filled squares) modeled with the SGF approach, and the Ge slabs (filled circles) with different slab thicknesses $L$ given in units of the lattice constant of bulk Ge, $a$.}
\label{fig:figure_BA2}
\end{figure}

To compare the electronic structures of the Ge(001) surface calculated with the slab and SGF models, we have done slab (SGF) calculations of the energy gap for the Ge(001) surface as a function of the thickness of the Ge slab (Ge surface region) adopted for modeling of the surface. In Fig.~\ref{fig:figure_BA4}, we show the local density of states (LDOS) calculated at the innermost region of the slab and of surface for two representative thicknesses, $L/a= 3, a$ and $L/a = 7$. One can see that the energy gap extracted from the LDOS decreases considerably when increasing the thickness from $L/a= 3$ to $L/a = 7$, while the energy gap in the innermost region of the slab remains essentially constant and matches the band gap of bulk Ge. In Fig.~\ref{fig:figure_BA2}, one can see that the energy gap value for the Ge(001) slab goes slowly to its asymptotic value (which coincides with the band gap of bulk Ge in this case) upon increasing the slab thickness. Contrary to the slab calculations, the energy gap of the Ge(001) surface modeled with the SGF approach is essentially constant across the system, as expected for a surface free from surface states, and does not depend on the value of $L$. The energy gap of the Ge(001) surface modeled with the SGF approach shows no further dependence on $L$ if the surface region thickness $L/a\geq 4$, whereas there exists a strong thickness dependence of the Ge(001) surface energy gap for the slab model. That suggests that the SGF model of the Ge surface accurately represents the bulk Ge states, eliminating any quantization effects, unlike the Ge slab model where quantization effects are sizable, even for $L/a=7$, see Fig.~\ref{fig:figure_BA2}.

\subsection{Si film on the Ge(001) surface} \label{sec:SiGe_interface}

In this section, we study a $\langle$001$\rangle$-oriented Si film interfaced with the Ge(001) surface. The main goal of this study is to show how the band alignment at the Ge(001)$\vert$Si interface can be calculated for different doping levels of the Ge substrate, using the SGF approach.

\begin{figure}
  \includegraphics[width=\linewidth]{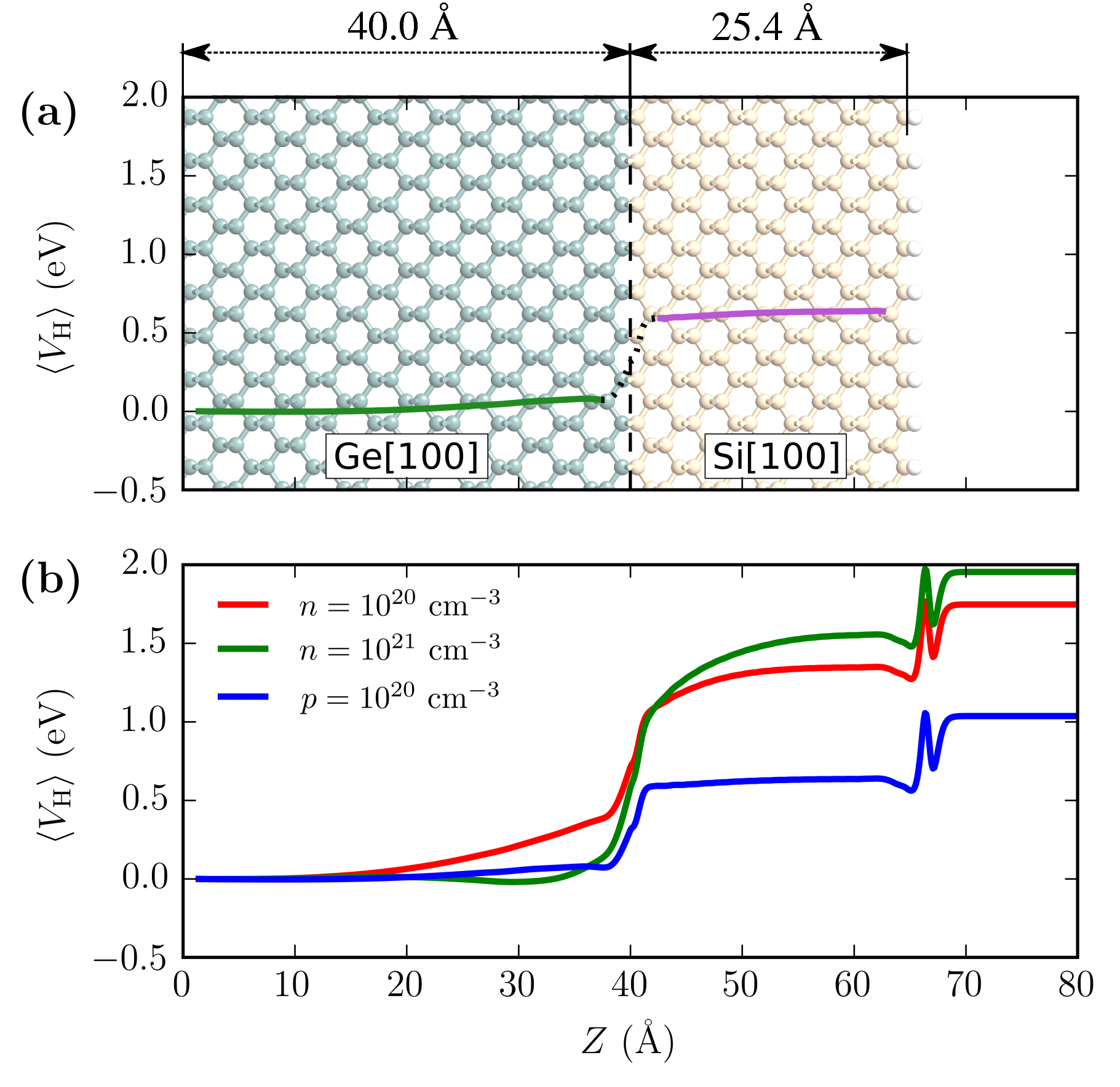}
  \caption{a) The Ge(001)$\vert$Si interface structure and the macroscopic in-plane averaged Hartree difference potential, $\langle \delta V_\mathrm{H} \rangle$, calculated for the Si film on the $p$-doped Ge substrate ($p = 10^{20} \mathrm{cm}^{-3}$). The green solid, black dotted and dark magenta solid lines indicate the regions of $\langle \delta V_\mathrm{H} \rangle$ corresponding to the Ge(001) surface, the Ge(001)$\vert$Si interface, and the Si film, respectively. Note that for visualization purposes the structure has been repeated in the $XY$-plane, while only a single lateral unit cell is used for the actual calculations. b) The macroscopic in-plane averaged Hartree difference potential, $\langle \delta V_\mathrm{H} \rangle$, along the $Z$-direction for the $p$-doped (blue), and $n$-doped (red, green) Ge substrate.}
\label{fig:ge_si_hpot}
\end{figure}

\begin{figure*}
  \includegraphics[width=0.9\textwidth]{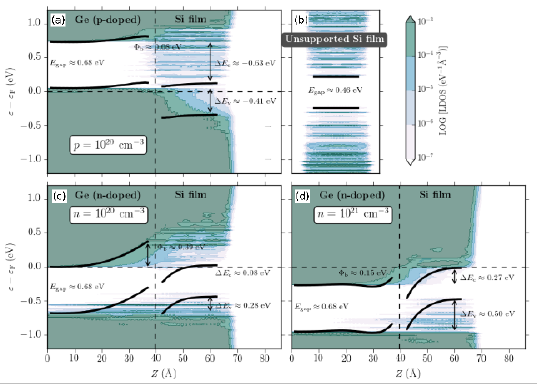}
  \caption{Band diagrams showing the density of states (DOS) across the Ge(001)$\vert$Si heterostructure for (a) $p$-doped Ge(001) substrate with $p = 10^{20} \mathrm{cm}^{-3}$, and $n$-doped Ge(001) substrate with (c) $n = 10^{20} \mathrm{cm}^{-3}$ and (d) $n = 10^{21} \mathrm{cm}^{-3}$. Notice that $\langle \delta V_\mathrm{H} \rangle$ in the Si region has been shifted to match the energy gap of the Si film. (b) Band diagram of the unsupported Si film (Si slab) showing the $\langle \delta V_\mathrm{H} \rangle$ superimposed on the DOS.}
\label{fig:PLDOS_30_atoms}
\end{figure*}

\paragraph*{\label{sec:MethodsBA2} Methods}

To keep the focus on application of the SGF methodology to the band alignment calculation rather than on understanding of the actual complex structure of the lattice-mismatched Ge(001)$\vert$Si interface, we have adopted a simple model to match a Si film on a Ge substrate, where the in-plane lattice parameter of the (minimal) lateral unit cell of the Si film is adjusted to that of the Ge(001) surface. This matching procedure gives rise to the lateral strain of 5.5 \% in the Si film. The Si film thickness is chosen to be 2.54 nm. The corresponding Ge(001)$\vert$Si heterostructure is illustrated in Fig.~\ref{fig:ge_si_hpot}a.

We have studied the Ge(001)$\vert$Si heterostructure for four different doping levels of the Ge(001) substrate, adopting the atomic compensation charge method for doping the semiconductor structure, see Refs.~\onlinecite{Stradi2016, Stradi2016_notes} for more details. We have used the SG15-Medium (High) combination of norm-conserving pseudopotential and LCAO basis set for silicon (germanium). All other computational settings are as for the Ge surface calculations in the previous section.  In addition, we have done ion relaxation for the top layers of the Ge(001) surface, as well as for the entire Si film in the heterostructure. The forces have been converged to a maximum value of 0.005 eV/\AA . The ion relaxation has been allowed in the out-of-plane $Z$-direction only, meaning that the Si film is still strained in the in-plane $X$ and $Y$-directions.

\paragraph*{\label{sec:ResultsBA2} Results}

The high strain in the supported Si film strongly reduces the energy gap of the corresponding Si slab from 1.28 eV to 0.46 eV as seen in Fig.~\ref{fig:PLDOS_30_atoms}b. Interfacing the Si film with the Ge(001) surface gives rise to a charge transfer from the Ge(001) to Si surface. Table~\ref{tab:si_ge} shows the charge ($Q_{\mathrm{Si}}$) induced in the Si film upon formation of the Ge(001)$\vert$Si heterostructure for three doping levels of the Ge(001) substrate. Note that $Q_{\mathrm{Si}}$ has been scaled with respect to its value at  $p = 10^{20} cm^{-3}$, $Q_{p-\mathrm{Si}}$. The charge transfer results in electron accumulation on the Si film. The electron accumulation further increases for increasingly larger $n$-doping of the Ge(001) substrate.

Figure~\ref{fig:ge_si_hpot}b shows the macroscopic in-plane averaged Hartree difference potential, $\langle \delta V_\mathrm{H} \rangle$,\cite{EDPnote,Baldereschi1988} for the different doping levels. One can see that $\langle \delta V_\mathrm{H} \rangle$ in the Si film goes upwards with respect to the potential in the bulk Ge region upon increasing the $n$-doping level in the Ge(001) substrate. This behavior of the electrostatic potential is due to the electron transfer from the Ge(001) to Si surface. The corresponding electric field that arises from the negative charge in the Si film penetrates into the Ge(001) substrate. To quantify the band alignment at the interface between the Si film and the Ge(001) substrate, we have calculated the DOS across the heterostructure for each doping level, see Fig.~\ref{fig:PLDOS_30_atoms}. In the case of $p$-doping, the Ge(001) surface is in the hole accumulation regime, and the charge transfer from the Ge(001) to Si surface results in a short screening length for $\langle \delta V_\mathrm{H} \rangle$, see Fig.~\ref{fig:PLDOS_30_atoms}a. When the Ge(001) substrate is $n$-doped, the Ge(001) surface is in the electron depletion regime that gives rise to a longer screening length (see Fig.~\ref{fig:PLDOS_30_atoms}c) compared to the case of $p$-doping for comparable magnitudes of the doping level. For a higher $n$-doping level, the screening length gets significantly reduced as seen in Fig.~\ref{fig:PLDOS_30_atoms}d, meaning that the space charge is confined in the Ge(001) near-surface region. Table~\ref{tab:si_ge} suggests that the charge redistribution across the entire heterostructure gives rise to a doping-dependent potential barrier, $\Phi_b$, at the Ge(001)$\vert$Si interface.

\begin{table}
\caption{\label{tab:si_ge} Band alignment parameters of the Ge(001)$\vert$Si interface for the three different doping levels of the Ge(001) substrate. $Q_{\mathrm{Si}}$ ($Q_{p\text{-Si}}$) is  is the induced charge in the ($p$-doped) Si film, where $Q_{p\text{-Si}}=0.49 \times 10^{12}$ $e/\mathrm{cm}^{2}$. $\Phi_b$ is the Schottky barrier at the Ge(001)$\vert$Si interface. $\Delta E_v$ ($\Delta E_c$) is the offset between the Si film and bulk Ge valence (conduction) band minima. Note that the $Q_{\mathrm{Si}}$ charge is calculated by integrating the electron difference density over the Si film.\cite{EDPnote}}
\begin{ruledtabular}
\begin{tabular}{lcccc}
  Doping level & $Q_{\mathrm{Si}}-Q_{p\text{-Si}}$ & $\Phi_b$ & $\Delta E_v$ & $\Delta E_c$ \\
  $\mathrm{cm}^{-3}$ & $e/\mathrm{cm}^2$ & eV & eV & eV  \\
\hline \cline{1-5} \\ [-2ex]
$p=10^{20}$ & $0$ & $0.08$ & $-0.41$ & $-0.63$ \\
$n=10^{20}$ & $-2.44 \times 10^{12}$ & $0.39$ &  $0.28$ &  $0.08$ \\
$n=10^{21}$ & $-6.70 \times 10^{12}$ & $0.15$ &  $0.50$ &  $0.27$ \\
%$p=10^{20}$ & $4.94 \times 10^{11}$ & 0.08 & -0.41 & -0.63 \\
%intrinsic  & $-1.4 \times 10^{11}$ & 0.34 & -0.06 & -0.30 \\
%$n=10^{20}$ & $-1.95 \times 10^{12}$ & 0.39 &  0.28 &  0.08 \\
%$n=10^{21}$ & $-6.21 \times 10^{12}$ & 0.15 &  0.50 &  0.27 \\
\end{tabular}
\end{ruledtabular}
\end{table}

Figure~\ref{fig:PLDOS_30_atoms} also shows a plot of $\langle \delta V_\mathrm{H} \rangle$ (overlaid on the DOS) that defines the actual edges of the bands as demonstrated in Ref.~\onlinecite{Stradi2016}. On the Ge side,  this is achieved by shifting $\langle \delta V_\mathrm{H} \rangle$ of an energy equal to the difference between the Fermi energy and the conduction band minimum (CBM) or the valence band maximum (VBM) in bulk Ge. In the silicon thin-film, $\langle \delta V_\mathrm{H} \rangle$ is shifted an energy equal to the difference between the Fermi energy and the CBM or the VBM in the corresponding silicon slab. Figure~\ref{fig:PLDOS_30_atoms} allows us to extract the band alignment parameters such as the interface potential, $\Phi_b$, which is given by the distance between the Ge CBM at the interface and in the bulk Ge region. The $\Phi_b$ potential acts as a barrier for the electron injection from the Ge(001) substrate into the Si film. Another band alignment parameter of relevance is the conduction (valence) band offset $\Delta E_c$ ($\Delta E_v$) that we define as the distance between the conduction band minimum (valence band maximum) in the bulk Ge region and the surface region of the Si film. A positive sign of the conduction band offset ($\Delta E_c>0$) indicates that there exists a potential barrier for the electrons propagating from the bulk Ge region to the Si film. The band alignment parameters extracted from the data shown in Fig.~\ref{fig:PLDOS_30_atoms} are listed in Table~\ref{tab:si_ge}.
For each of the three heterostructures with different doping levels, $\Delta E_v$ and $\Delta E_c$ have the same sign, meaning that the band alignment is of type II with staggered Ge and Si gaps. However, if the conduction and valence band offsets are defined right at the Ge(001)$\vert$Si interface, the $p$-doped and intrinsic heterostructures have a type III broken gap. We notice that applying Anderson's electron affinity rule would result in a qualitatively different band diagram for the Ge(001)$\vert$Si heterostructure compared to that obtained from the present first-principles study. That suggests that using this empirical rule might not reliably predict the band alignment in complex heterostructures where microscopic details of the interfaces between dissimilar semiconducting materials matter.

Figure~\ref{fig:PLDOS_30_atoms} also suggests that there exist Ge(001) states that penetrate into the Si film, and this state penetration is related to one of the mechanisms responsible for the electron donation to the Si film, in agreement with earlier predictions for semiconductor heterojunctions.\cite{tersoff1984theory} This is particularly evident for the highly $n$-doped Ge(001) substrate as shown in Fig.~\ref{fig:PLDOS_30_atoms}d, where we see that the $\langle \delta V_\mathrm{H} \rangle$ potential in the Si film virtually follows the DOS penetration profile related to the conduction band states of the near Ge(001) surface region.

Using Fig.~\ref{fig:PLDOS_30_atoms}, one can conclude that some Si states also penetrate into the Ge(001) substrate. In particular, for the highly $n$-doped Ge(001) substrate this gives rise to a non-monotonic behavior of the $\langle \delta V_\mathrm{H} \rangle$ potential, which has a minimum near the interface. Note that there exist no midgap energy levels at the surface of the Si film as seen in Fig.~\ref{fig:PLDOS_30_atoms}, confirming that hydrogen passivation of the Si(001) surface has efficiently removed all the surface point defects related to the Si dangling bonds. Similarly, we do not find any localized interface states at the Ge(001)$\vert$Si interface. All the states at the interface arise from penetration of either Ge substrate or Si film states across the interface.

In conclusion, the present study demonstrated that the SGF approach provides an insightful, accurate, computationally efficient way for calculation and analysis of complex semiconductor heterostructures at the microscopic level within the framework of DFT. We have shown that the SGF approach is superior compared to the commonly-used slab approach as it accounts for bulk states of semiconductor substrates in an exact manner, unlike the slab approach that suffers from finite-size effects.

\section{\label{sec:SS}Surface states}
Electronic surface states are notoriously difficult to describe using
the slab method, as equivalent states localized on both surfaces of the slab will interact
strongly if the slab is not thick enough.  \textit{A-posteriori} corrections\cite{Berland2012} are then needed
to decouple the surface states and to correctly model the field dependence of the surface state properties.
In this section, we show that the surface states are
naturally taken into account within the framework of the SGF method,
which deals with a single surface only, and external fields can be applied by shifting the potential near the surface in the
vacuum in a simple manner.

We focus here on studying the Shockley surface state that is present at the center of the Brillouin zone on the (111) surfaces of noble metals,\cite{Shockley1939}
and the topologically-protected surface states that are present at the surface of
topological insulators (TIs).\cite{Zhang2010,Xia2010}
As an example of a Shockley-type surface state, we consider the Ag(111) surface,
for which accurate experimental data from photoemission spectroscopy (PES)
and scanning tunneling spectroscopy (STS) are available.
It has also been demonstrated that external electric fields
can alter the surface state and change its overall properties.\cite{Limot2003,Kroger2004}
As a prototypical TI surface, we consider a Se-terminated Bi$_2$Se$_3$(111) surface.
A previous work has also adopted a SGF-type approach
to describe the formation of surface states on the Bi$_2$Se$_3$(111) surface,
but that approach was based on a parametrized effective Hamiltonian.\cite{Zhang2010,Zhang2010b}
In the following, we give a first-principles, atomistic description of the surface states that is not based on
any adjustable parameters.

For both the Ag(111) and Bi$_2$Se$_3$(111) surface, the surface states are identified
in the surface band structure, which has been described with the density of states (DOS) calculated
along the $M \to \Gamma \to K$ $k$-path in the 2D Brillouin zone (BZ) of the surface.
\subsection{Ag(111) Shockley surface state}
\paragraph*{\label{sec:SSMethods}Methods}

We have done the ATK-SGF calculations using a surface region comprised of 27 atomic monolayers and a vacuum layer with a thickness of 20~\AA.
This large number of Ag(111) monolayers is used to increase the contribution of the bulk Ag states to the electronic
structure of the surface region projected onto the 2D BZ of the (111) surface.
We notice that the Shockley surface state is highly-localized at the surface, and it can be accurately described by
using just 7 atomic monolayers in the surface region. It means that we could adopt, in principle, a smaller surface region, adding the DOS of bulk Ag to the SGF-calculated DOS of the surface region.

\begin{figure}
\includegraphics[width=\linewidth]{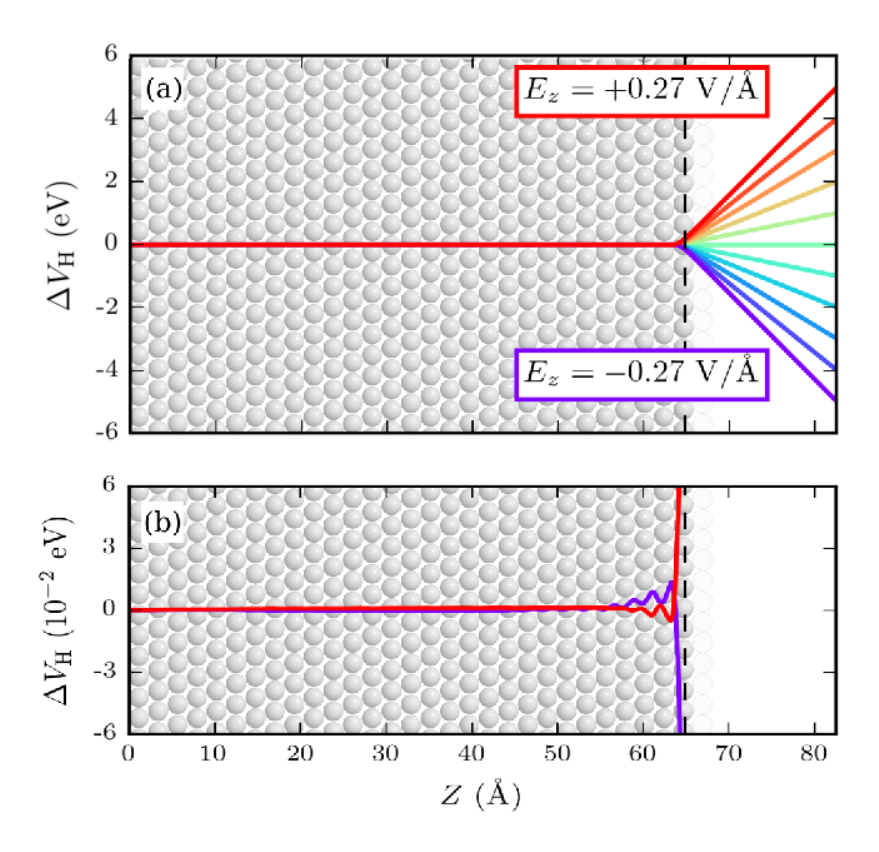}
\caption{
(a) Field-induced difference in the macroscopic in-plane averaged Hartree potential in the Ag(111)
surface region, $\Delta V_\mathrm{H}$, plotted in the out-of-surface-plane
($Z$) direction, for different external electric fields,
$E_z$.
%ranging
%from $E = - 0.27\ \mathrm{V}/\mathrm{\AA}$ (violet line) to
%$E = + 0.27\ \mathrm{V}/\mathrm{\AA}$ (red line).
The interface between the semi-infinite bulk Ag region and surface region is at $Z = 0\ \mathrm{\AA}$.
The black dashed vertical line at $Z_\mathrm{surf} = 64.68\ \mathrm{\AA}$
indicates the position of the Ag(111) top monolayer.
Ag atoms are shown as gray spheres.
(b) $100\times$zoom of the inset (a) for $E_z = \pm 0.27\ \mathrm{V}/\mathrm{\AA}$.
}
\label{fig:figure_SS1}
\end{figure}

The surface 2D BZ has been sampled using a 21$\times$21 $k$-point grid.
The corresponding 3D BZ in the bulk electrode has been sampled
using 21$\times$21$\times$201 $k$-points. Following the procedure described in Ref.~\onlinecite{GarciaGil2009}, we have included a layer of
ghost atoms above the top monolayer of the surface to accurately describe the decay of the surface electron density
into the vacuum.
The SGF surface calculations have been done for different external electric fields applied perpendicularly to the surface plane,
ranging from $E_z=-0.27$~$\mathrm{V}/\mathrm{\AA}$ to $E_z=+0.27$~$\mathrm{V}/\mathrm{\AA}$,
at a regular step of $\Delta E_z=0.054$~$\mathrm{V}/\mathrm{\AA}$.
In the SGF method, an electric field is imposed with the Dirichlet boundary condition by shifting the
electrostatic potential value in the vacuum, while keeping the chemical potential of the semi-infinite bulk region unchanged.
Note that this procedure resembles an experimental measurement in which the surface is exposed to an external field
generated by a scanning tunneling microscopy tip.\cite{Limot2003}
The Ag(111) surface structure shown in Fig.~\ref{fig:figure_SS1} has been built using the DFT-PBE calculated
lattice constant of bulk Ag ($a_\mathrm{Ag} = 4.086~\mathrm{\AA}$). Subsequently, we have done ion relaxation for the top surface layers.
More information on the computational details of the ATK-SGF calculations can be found in Sec.~\ref{sec:compdetails}.
\paragraph*{\label{sec:SSResults}Results}
\begin{figure}
\includegraphics[width=\linewidth]{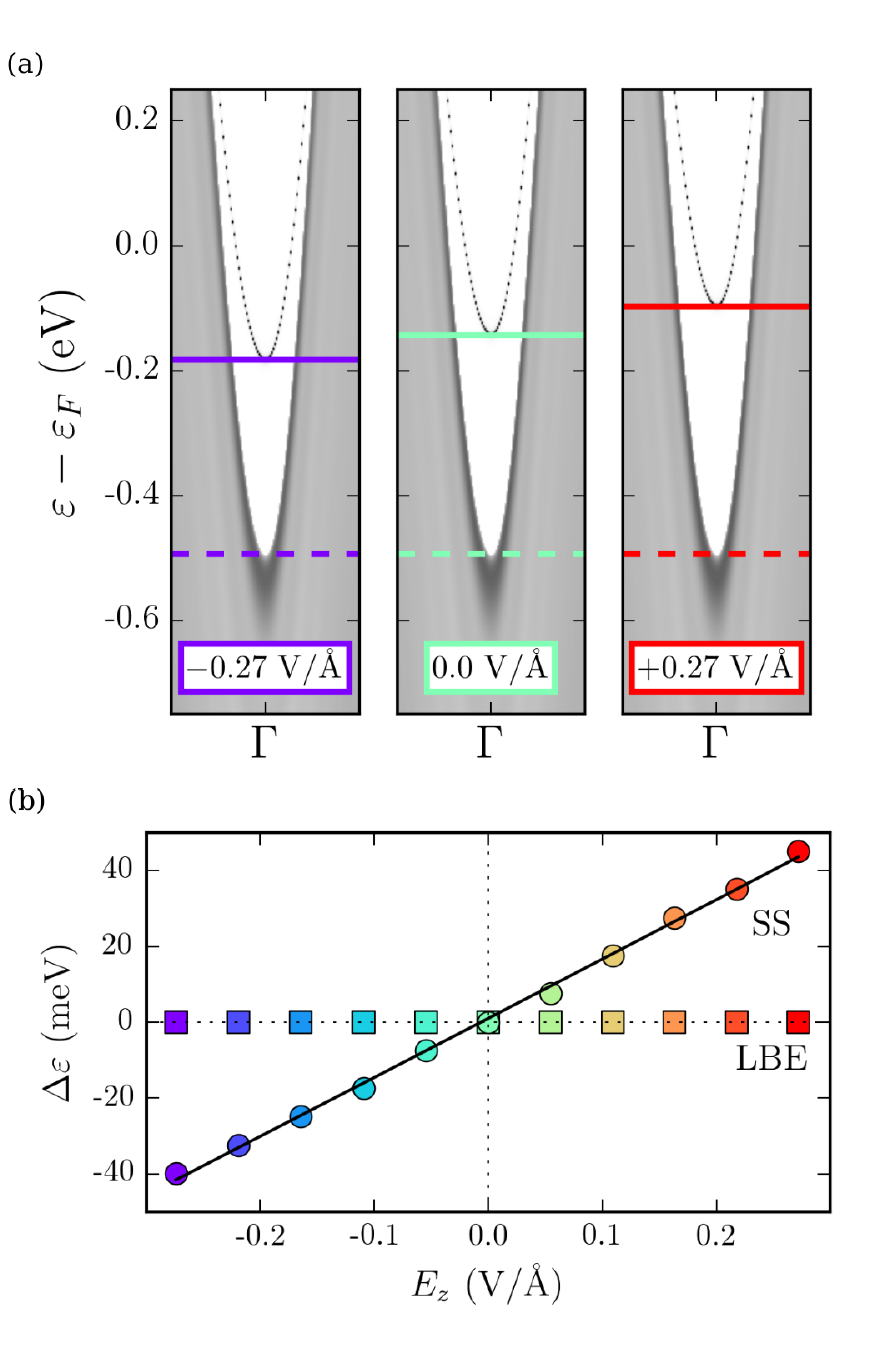}
\caption{
(a) Band structure of the Ag(111) surface along the
$\mathrm{M} \to \Gamma \to \mathrm{K}$ $k$-path of the 2D Brillouin zone
in the vicinity of the Fermi energy and close to the $\Gamma$ point,
for applied fields of $-0.27\ \mathrm{V}/\mathrm{\AA}$ (left inset),
$0.0\ \mathrm{V}/\mathrm{\AA}$ (center inset),
and $+0.27\ \mathrm{V}/\mathrm{\AA}$ (right inset).
The horizontal solid (dashed) lines indicate the
lowest energy of the surface states (highest occupied bulk states) at the $\Gamma$ point.
(b) Stark energy shift ($\Delta \varepsilon$) of the Ag(111) surface state against the $E_z$-electric field applied.
Filled {\it circles} and black {\it solid line} correspond to the surface state (SS) and a linear fit of the calculated data, respectively. The filled {\it squares} correspond to lower band edge (LBE).}
\label{fig:figure_SS2}
\end{figure}
Figure~\ref{fig:figure_SS1} shows the difference ($\Delta V_\mathrm{H}$) in the Hartree potential
induced by the external electric field,
\begin{equation}
\Delta V_\mathrm{H}=V_\mathrm{H}-V_\mathrm{H}^0,
\end{equation}
where $V_\mathrm{H}^0$ is the Hartree potential at zero field.
Figure~\ref{fig:figure_SS1} suggests that applying the external field induces
a perturbation of the Hartree potential that is far beyond the Ag(111) topmost monolayer,
located at $Z_\mathrm{surf} = 68.64\ \mathrm{\AA}$.
For the electric field magnitude of $E_z = \pm 0.27\ \mathrm{V}/\mathrm{\AA}$, the oscillations of the $\Delta V_\mathrm{H}$ potential at $Z < Z_\mathrm{surf}$, which are clearly seen in Fig.~\ref{fig:figure_SS1}b, indicate that
the field-induced perturbation of the surface electronic structure
is completely screened after the 7th innermost Ag(111) monolayer only,
with the screening being somewhat more efficient for positive than for negative biases.
%This implies that, disregarding quantum confinement effects,
%a slab of at least 14 layers is needed
%in order to correctly describe the screening response of the surface
%in this range of electric fields,
%while the SGF method needs a central region of only 7 layers.
%Also note in Fig.~\ref{fig:figure_SS1} that
%$\Delta V_\mathrm{H} = 0\ \mathrm{eV}$ between the screening region
%and the electrode at $Z = 0\ \mathrm{\AA}$,
%indicating that the surface electronic structure is seamlessly matched
%to that of the bulk reservoir.

The 2D surface electronic band structure of the Ag(111) surface is shown in Fig.~\ref{fig:figure_SS2}a
for $E_z = 0\ \mathrm{V}/\mathrm{\AA}$ (green) and $E_z = \pm 0.27\ \mathrm{V}/\mathrm{\AA}$
(violet and red).
The bottom of the surface state band (indicated by solid lines)
is located at the $\Gamma$ point, above the highest occupied bulk state
(indicated by dashed lines).
At zero field, the energy at the bottom of the surface state band is
$\varepsilon - \varepsilon_\mathrm{F} = -147$~meV,
in good agreement with the value of $-120\pm1$~meV
%$\varepsilon - \varepsilon_\mathrm{F} = -120\pm1$~meV
obtained from PES measurements.\cite{Kevan1989}
A fit of the parabolic dispersion
of the surface state band using the free-electron gas model,
$\varepsilon = (\hbar k)^2 / 2 m^*$,
results in a value for the electron effective mass of $m^* = 0.306\ m_e$,
in close agreement with the value $m^* = 0.31\pm0.01\ m_e$
measured with STS.\cite{Garnica2016}

Applying an electric field gives rise to a linear Stark shift ($\Delta \varepsilon$)
of the surface state energy, which follows the sign of the applied field.
This behavior is clearly seen in Fig.~\ref{fig:figure_SS2}a,
and is consistent with several experimental\cite{Limot2003,Kroger2004}
and theoretical reports.\cite{Berland2012}
Figure~\ref{fig:figure_SS2}b shows how the Stark shift computed for the Ag(111) surface states
depends on the external electric field. A linear fit to the $\Delta \varepsilon$ \textit{vs.} $E_{z}$ data
in Fig.~\ref{fig:figure_SS2}b yields a slope of $8 \times 10^{-3}\ e\mathrm{\cdot \AA}$.
It is evident that the field alters the dispersion of the surface state bands, resulting in a variation of
the electron effective mass from $m^* = 0.301\ m_e$  to $m^* = 0.314\ m_e$
corresponding to $E_{z}=-0.27$ and $+0.27\ \mathrm{V/\AA}$, respectively. This is in agreement
with the results previously-reported for the Cu(111) surface state.\cite{Berland2012}

Strikingly, the variation of the Stark shift, $\Delta \varepsilon$, is linear with respect to the $E_z$
even in the limit of a vanishing field, when the shift calculated with the slab model would exhibit an avoided crossing behavior
as a result of the interaction between the surface states that are related to the two surfaces of the slab.\cite{Berland2012}
Figure~\ref{fig:figure_SS2}b also shows that the position of the lower
band edge (LBE) of the bulk bands remains fixed in the SGF-calculated band structure of the Ag(111) surface, while $E_z$ changes.
We notice that this physically-correct behavior is not captured by the slab model, as the Ag states of the thin slab structure are not pinned to the true bulk Ag states, and therefore the corresponding bulk-like bands of the slab can be shifted by the applied electric field.
\subsection{Bi$_2$Se$_3$(111) topologically-protected surface state}
\paragraph*{Methods}
To study the topologically-protected surface states on the Bi$_2$Se$_3$(111) surface, we have constructed the Bi$_2$Se$_3$(111) surface structure, using a fully-relaxed bulk Bi$_2$Se$_3$ unit cell, where the forces and stress were converged to 0.05~eV/$\mathrm{\AA}$ and 1~GPa, respectively.
The surface region comprises 37 atomic monolayers, corresponding to 7.6 quintuple layers (QLs). The principal layer of the bulk region consists of 3 QLs. For the sake of simplicity, we have not done ion relaxation of the surface.
A non-collinear spin formalism including spin-orbit coupling
has been employed in all the ATK-SGF calculations of the topologically-protected surface states.\cite{Chang2015}
The 2D (3D) BZ of the surface (bulk) region has been sampled with a 9$\times$9 (9$\times$9$\times$201) $k$-point grid,
and the broadening of the Fermi--Dirac distribution for calculating the electron occupation has been set to a rather small value of $\sim0.004$~eV.

\paragraph*{Results}
\begin{figure}
\includegraphics[width=\linewidth]{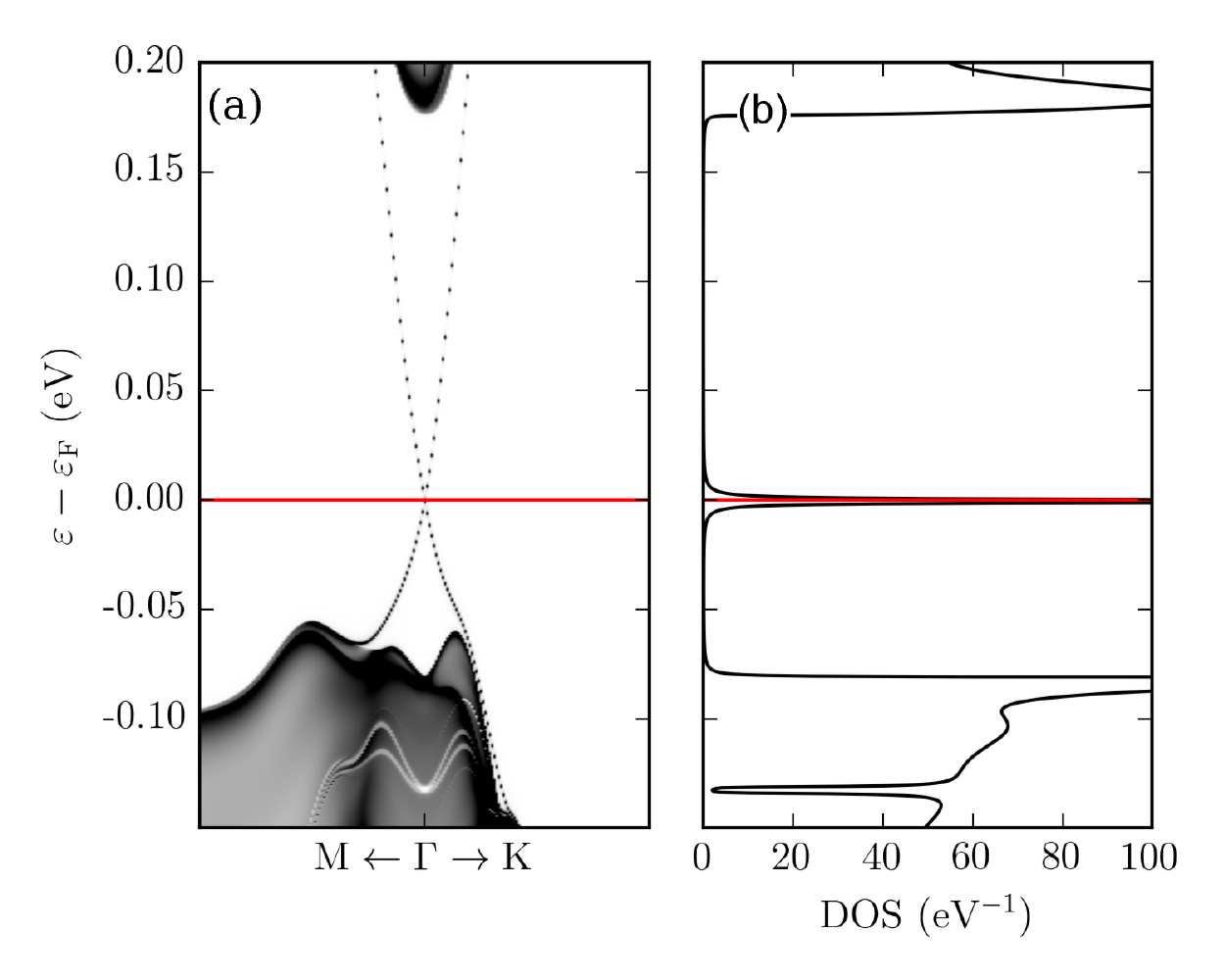}
\caption{(a) Band structure of the Se-terminated Bi$_2$Se$_3$(111) surface
along the $M \to \Gamma \to K$ $k$-path in the 2D Brillouin zone
in the vicinity of the Fermi energy, close to the $\Gamma$ point.
(b) Density of states at the $\Gamma$ point.
The red horizontal lines in both insets indicate the position
of the Fermi energy.}
\label{fig:figure_SS3}
\end{figure}
In Fig.~\ref{fig:figure_SS3}a, one can see the electronic band structure
of the Bi$_2$Se$_3$(111) surface, calculated with the ATK-SGF method.
This figure suggests that there exist two topologically-protected surface states
inside the electronic gap of bulk Bi$_2$Se$_3$, as the two surface states cross
at the Fermi energy (the Dirac point), around which the dispersion is essentially linear.
This is in agreement with previous work where either the slab\cite{Chang2015}
or SGF approach\cite{Zhang2010,Zhang2010b} have been adopted to study the topologically-protected surface states.

By examining the the ATK-SGF calculated surface DOS at the $\Gamma$-point
(see Fig.~\ref{fig:figure_SS3}b),
we find that the electronic energy gap of the bulk material is of 250~meV,
and the conduction band minimum is at 170~meV
above the Dirac point associated with the surface states,
in good agreement with the values reported in earlier angle-resolved
PES measurements on Bi$_2$Se$_3$ single crystals.\cite{Xia2010}
Importantly, a single narrow peak is present at the Fermi energy,
which is related to the spin-degenerate state arising from
the intersection between the two spin-locked surface states.
The peak has a Lorentzian shape with a width given by the actual numerical value of the infinitesimal, $\delta$,
used for computing the Green's function of the surface region in Eq.~\eqref{eq:negf_2}.
This degeneracy between the two surface bands arises naturally within the framework of the SGF formalism,
whereas in finite-size slab models of the topological insulator surfaces, the interaction between evanescent states
localized at the two surfaces of the slab leads to an unphysical energy gap opening
that is inversely proportional to the slab thickness.\cite{Yazyev2012} This allows us to conclude that the SGF method provides an accurate description of the topologically-protected surface states, compared to the slab method.

\section{\label{sec:surface_chemistry}Surface chemistry in external electrostatic fields}
The properties of adsorbed species at electrochemical metal--solution
interfaces depend on the applied electrode potential
and hence the electric field.
Several theoretical works have considered the response of chemisorption
binding energies and vibrational frequencies to a potential bias,
using either slab calculations or metallic clusters.\cite{neugebauer1993,wasileski2001}
In particular, Bonnet and co-workers have recently used the slab model
in combination with the effective screening medium \cite{otani2006} method
to investigate the vibrational response of carbon monoxide
on a platinum electrode from first principles.\cite{bonnet2014}
Other works have focused on the very high fields needed to rip atoms out of the surface during field emission processes.\cite{Gomer1994,sanchez2004}

A slab is a confined system in the out-of-plane direction, so any charging of adsorbates on the slab surface must be counter-balanced by an opposite charge in the slab, altering the electron chemical potential of the finite-size slab system. We notice that no change of the chemical potential would take place in a truly semi-infinite surface system. In the SGF approach, the chemical potential of the surface with adsorbates is fixed by an infinite reservoir of electrons (bulk region) coupled to the surface region. The electrons are allowed to be transferred between the surface and bulk regions in a fully self-consistent manner. The adsorbates may therefore be charged with the charges that originate from the bulk region without altering the chemical potential of the surface system, unlike the slab system.

We here consider atomic iodine adsorbed on the Pt(111) surface,\cite{Tkatchenko2005,Schardt1989} which is a system of relevance for dye-sensitized solar cells.\cite{Zhang2013}
We show that the SGF method is a natural choice for studying the chemical properties of adsorbates on the surface in an external electrostatic field, as it allows charging of the iodine atom from the electron reservoir, instead of the limited electron supply in a slab system.
\begin{figure}
\includegraphics[width=\columnwidth]{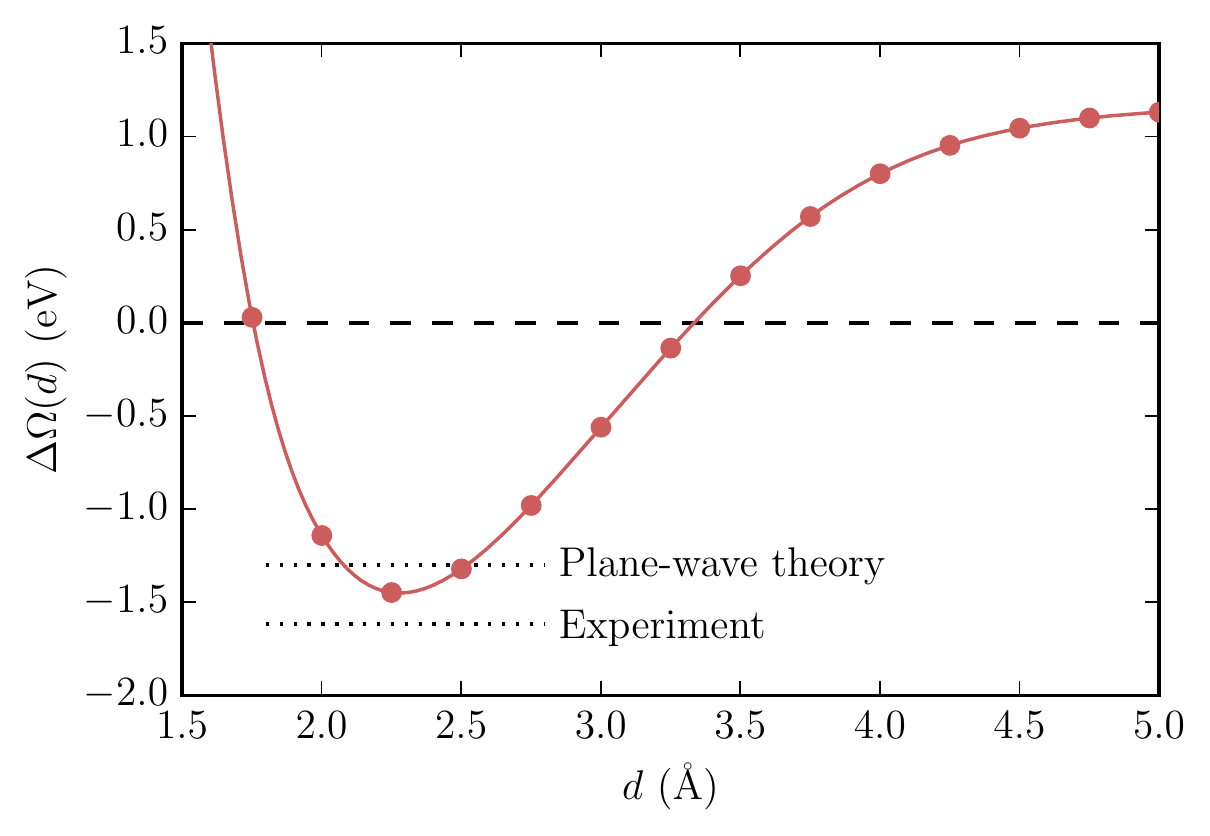}
\caption{Zero-field potential-energy curve for atomic iodine
adsorbed on the Pt(111) surface,
calculated using the SGF method.
Cubic interpolation of the PEC near the minimum yields an adsorption energy
of $-1.45$~eV at an equilibrium I--Pt(111) separation distance ($d_{0}$) of
2.28~\AA.
The measured and plane-wave DFT-calculated
adsorption energies (dotted lines) are taken from Ref.~\onlinecite{wellendorff2015}.}
\label{fig:figure_chem1}
\end{figure}
\begin{figure}
\includegraphics[width=\columnwidth]{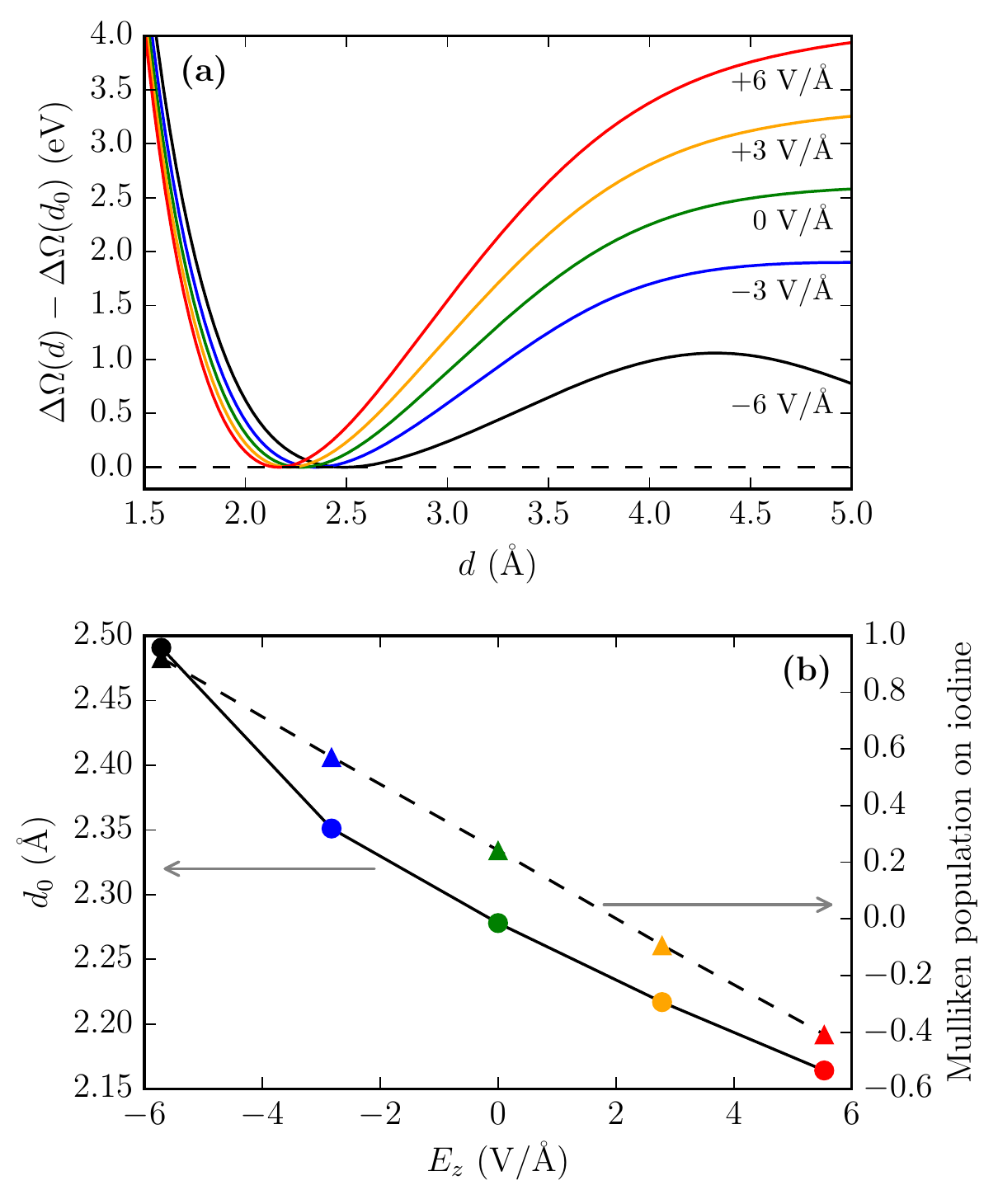}
\caption{(a) Finite-field potential-energy curves for atomic iodine
adsorbed on the Pt(111) surface, for electrostatic fields ranging from
$+6$~V/\AA\ (red) over zero field (green) to $-6$~V/\AA\ (black).
For all the curves, zero energy is chosen at the equilibrium position
of the adsorbate.
(b) Equilibrium I--Pt(111) separation distance, $d_{0}$, (solid line)
and Mulliken population (dashed line) on the iodine atom
as a function of the applied field.
Positive Mulliken population means accumulation of electrons on the iodine atom.}
\label{fig:figure_chem2}
\end{figure}
\paragraph*{Methods}
We have constructed a $2\! \times\! 2$ Pt(111) surface, using a crystal structure of
bulk Pt with the DFT-optimized lattice parameter, $a_\mathrm{Pt}$ = 3.956~\AA. We have adopted 3 atomic (111) monolayers for the principal layer of the bulk region, and 9 atomic (111) monolayers for the central region. The vacuum thickness has been set to 20~\AA . The Neumann boundary condition has been imposed in the vacuum region. The 2D Brillouin zone has been sampled using a $6\! \times\! 6$ $k$-point grid. The top 6 monolayers of the Pt(111) surface have been relaxed within the framework of the DFT approach, see Sec.~\ref{sec:compdetails} for more computational details.
For surface calculations with adsorbates, atomic iodine has been placed in the fcc hollow site, and the iodine atom and top 6 monolayers of the Pt(111) surface have been relaxed, converging forces to 0.05~eV/\AA.
%The resulting Pt--I bond length was 2.79~\AA\ to all three Pt atoms in the
%three-fold coordinated site.

To calculate the equilibrium separation distance between the iodine atoms in a single I$_2$ molecule, we have adopted a large unit cell with the sufficiently-thick vacuum padding around the molecule to avoid iteration between the repeating images. $\Gamma$-only $k$-points sampling and 4 meV broadening of the Fermi-Dirac distribution are used for this calculation, yielding an I--I equilibrium bond length of 2.73~\AA.

The potential-energy profile for the interaction of a single iodine atom with the Pt(111) surface has been calculated by displacing the I atom away from its equilibrium position on the surface along the surface normal in steps of $\delta d = 0.25$~\AA. For a given Pt(111)--I separation distance $d$, the energy of the system has been calculated by using the grand canonical potential, as defined for an open system coupled to an electron reservoir,
\begin{equation}
\Omega[\rho] = E[\rho]- e \, \delta n_\mathrm{bulk}\, \mu_\mathrm{bulk},
\end{equation}
where $E$ is the total energy of the surface region, $\rho$ is the electronic density in the surface region, and $\delta n_\mathrm{bulk}$ is the number of electrons exchanged with the the electron reservoir with chemical potential $\mu_\mathrm{bulk}$. The adsorption energy $\Delta \Omega$ is then evaluated as:
\begin{align}
\Delta \Omega(d) = & \Omega_\text{I/Pt(111)}(d) - \Omega_\text{Pt(111)} \nonumber \\
& - \tfrac{1}{2}E_{\text{I}_2} - \Delta_\text{CP}(d),
\end{align}
where $\Omega_\text{I/Pt(111)}$ and $\Omega_\text{Pt(111)}$ are the grand canonical potentials of the Pt(111) system with and without adsorbate, respectively. $E_{\text{I}_2}$ is the total energy of a I$_{2}$ molecule, which is equivalent to $\Omega_{\text{I}_2}$ since an isolated molecule cannot exchange particles with a reservoir. The counterpoise (CP) correction $\Delta_\text{CP}(d)$ is similar in spirit to the standard Boys-Bernardi CP correction to account for the basis set superposition error,\cite{counterpoise1970}
\begin{align}
\Delta_\text{CP}(d) = & \big(\Omega_\text{I*/Pt(111)}(d) - \Omega_\text{Pt(111)}\big) \nonumber \\
& + \big(E_\text{I/Pt(111)*}(d) - E_\text{I} \big) \nonumber \\
& + \tfrac{1}{2}\left(E_{\text{I}_2}^{*} - E_{\text{I}_2}\right), 
\end{align}
where $E_{\text{I}_2}^{*}$ is the total energy of a fictitious I$_{2}$ molecule in which one of the two iodine atoms is assumed to be a ghost atom, $\Omega_\text{I*/Pt(111)}$ is the grand canonical potential of the I/Pt(111) surface in which the iodine atom is treated as a ghost atom, and $E_\text{I/Pt(111)*}$ is the total energy of the corresponding I/Pt(111) slab in which the platinum atoms are treated as ghost atoms.

Field-dependent potential profiles have been obtained by imposing the Dirichlet boundary condition with different electrostatic potential values in the vacuum region, corresponding to external electric fields in the range from $-6$ to $+6$ V/\AA . We notice that, due to the use of a LCAO basis set, the present approach is not suitable for describing field-emission processes, in which electrons are moved from the surface to the vacuum, due to an applied electrical field.\cite{GarciaLekue2013}
\paragraph*{Results}

Figure~\ref{fig:figure_chem1} shows the potential profile
calculated using the DFT-SGF method for an iodine atom interacting with the Pt(111) surface.
The equilibrium adsorption energy obtained from this profile agrees well with the measured and plane-wave DFT-calculated energies.\cite{wellendorff2015}
The Pt(111)--I separation distance, $d$, is defined with respect to the top monolayer of the Pt(111) surface,
and the equilibrium separation distance, $d_{0}$, obtained with the DFT-SGF method is of 2.28~\AA.

Figure~\ref{fig:figure_chem2}a suggests that applying an external electrostatic field has a significant impact on
the SGF-calculated potential profile for fields in the range from $-6$~V/\AA\ to $+6$~V/\AA.
The potential-energy profile in the vacuum region is pushed down for increasingly large negative fields, lowering the energy barrier for desorption, whereas positive fields have the opposite effect. This behavior is opposite compared to that found for field-induced desorption of Na\citep{neugebauer1993} and Al\citep{sanchez2004} adatoms on Al(111), and can be ascribed to the propensity of the iodine adatom to form a stable anion, rather than a stable cation, due to its halogenic character. The equilibrium I--Pt(111) separation distance, $d_{0}$, is also affected by the electrostatic field change, as illustrated in Fig.~\ref{fig:figure_chem2}b.
In this figure, the Mulliken charge on the iodine atom is shown as function of the applied field strength.
As the field turns more negative, electron charge accumulates on the iodine atom, and the I--Pt(111) separation distance increases due to the larger anionic character of the adatom. As the I--Pt(111) distance is increased, the Mulliken population on the iodine atom remains essentially constant for negative applied fields, whereas it becomes increasingly more negative for positive values of the applied field,  reaching values of -0.37 $e^-$ ($E_z = +3\ \mathrm{V}$) and -0.85 $e^-$ ($E_z = +6\ \mathrm{V}$) at $d = 5\ \mathrm{\AA}$. In conclusion, we notice that this charge accumulation on the iodine atom does not require a corresponding charge of opposite sign in the near-surface region, as it is taken from the semi-infinite bulk region instead. That shows a crucial difference between the traditional slab and Green's-function approaches for surface chemistry calculations.

\section{\label{sec:Conclusions}Conclusions}
In this work, we presented the state-of-the-art implementation of the
Green's function formalism\citep{inglesfield1988surface,maclaren1989layer,skriver1991self,kudrnovsky1992self,szunyogh1994self,ishida2001surface,papior2017improvements} for accurate first-principles simulations of surfaces within the framework of density functional theory. Unlike the slab model that is traditionally used in computational surface science, the Green's-function approach allowed us to model the surface as a truly semi-infinite system by coupling a surface region to an electron reservoir. We were able to do first-principles calculations of surface systems that are free from the drawbacks present in the slab calculations, which are affected by finite-size effects. Furthermore, the computational cost of the Green's-function based surface calculations was shown to have a linear scaling with respect to the length of the surface region. For large systems, it provides a better alternative to the slab calculations that have a cubic scaling with respect to the slab thickness.

Using the Green's-function approach was shown to improve the accuracy of both quantitative and qualitative description of surface properties that are notoriously difficult to address using the slab approach, including metal work functions, surface states of metals and topological insulators, and energy gaps of semiconductor surfaces. We demonstrated the actual advantages of using Green's functions for several advanced physics and chemistry studies of surfaces. The adopted approach allowed us to accurately calculate the work functions of several transition metal surfaces. We found that the first-principles Green's-function approach combined with the analysis of physical properties based on the projected density of states and Hartree difference potential makes possible to quantitatively determine the band diagram across semiconductor heterostructures such as an ultra-thin Si film on an intrinsic or doped Ge substrate in atomistic simulations.  We found that it is crucial to adopt the surface Green's-function method for correct description of topologically-protected states in the Bi$_2$Se$_3$ topological insulator, as well as the effect of an external electric field on the surface state of the Ag(111) surface. The charge transfer effects for metal surfaces with adsorbates such as iodine atoms on the Pt(111) surface, turned out to be naturally captured within the framework of the Green's-function formalism that allows describing the surface structures coupled to an electron reservoir.

In conclusion, the present results suggested that the Green's-function approach to surface calculations is a superior tool compared to more traditional approaches to surface modeling. Given the demonstrated advantages of this approach, in this work we showed how one may increase the accuracy of DFT-based surface calculations, and how the applicability of first-principle, atomistic modeling can be extended towards challenging problems in surface science.
\bibliography{biblio}

\begin{thebibliography}{101}
\bibitem [{EDP()}]{EDPnote}%
  \BibitemOpen
  \href@noop {} {}\bibinfo {note} {The electrostatic difference potential is
  defined as $\delta V_\text{E} = V_\text{E}^\text{scf} -
  V_\text{E}^\text{sp}$, where $V_\text{E}^\text{scf}$ is the electrostatic
  potential of the self-consistent valence charge density and
  $V_\text{E}^\text{sp}$ the electrostatic potential from a superposition of
  atomic valence densities, $n^\text{sp}$. Similarly, we define the Hartree
  difference potential as $\delta V_\text{H} = V_\text{H}^\text{scf} -
  V_\text{H}^\text{sp} = -e\delta V_\text{E}$, where $e$ is the elementary
  charge ($e>0$), and the electron difference density as $\delta n =
  n^\text{scf} - n^\text{sp}$}\BibitemShut {NoStop}%
\end{thebibliography}
%\section{Appendix} \label{app:pps}
\begin{center}
\appendix{\bf APPENDIX}
\end{center}

\subsection{Pseudopotentials and basis sets}

The accuracy of the DFT calculations based on the pseudopotential LCAO approach depends on the choice
of pseudopotentials and basis sets. We employ norm-conserving pseudopotentials
in the Kleinman--Bylander form.\cite{Kleinman1982}
The basis functions are atom-centered orbitals constructed by solving
the Schr\"{o}dinger equation for a single atom in a confinement
potential.\cite{Soler2002,Blum2009}

In the ATK-2016 version, we have implemented high-accuracy pseudopotentials
and localized basis sets for all elements up to $Z=83$ (Bi), excluding lanthanides.
We have used the SG15 suite of optimized norm-conserving Vanderbilt pseudpotentials
from Ref.~\onlinecite{Schlipf2015}. For a number of chemical elements, we have improved the
pseudopotential quality by adding a nonlinear core correction.\cite{Louie1982}
Both scalar-relativistic and fully relativistic versions of all the pseudopotentials
are available in the ATK software package. The fully relativistic pseudopotentials allow for DFT calculations with spin-orbit
coupling included.\cite{Theurich2001}

To construct high-accuracy LCAO basis sets, we have first taken
a large set of pseudo-atomic orbitals similar to the ``tight tier 2''
basis sets used in the FHI-aims package.\cite{Blum2009} These
basis sets typically have 5 orbitals per pseudopotential valence electron,
a range of 5~\AA\ for all orbitals, and include angular momentum channels
up to $l=5$. We find that such a large LCAO basis set gives essentially
the same computational accuracy as fully-converged plane-wave calculations.
We have then reduced the range of the orbitals by requiring that the overlap of the contracted wave function must change less than 0.1 \% with the original wave function. Such a reduction of the orbital range decreases the number of matrix elements that needs to be evaluated, and at the same time this does not alter the accuracy of LCAO calculations. In the following, this LCAO basis set will be called {\it Ultra}.

From the Ultra basis set we generate two reduced basis sets, {\it High} and {\it Medium}.
The High basis set is generated by
reducing the number of basis set orbitals such that the DFT total energy
of suitably chosen test systems does not change by more than 1 meV (per atom).
For each element, the test set consists of the element in its experimental
(300~K) bulk structure at different lattice constants (that allows for computing
the $\Delta$-value\cite{Lejaeghere2014,DeltaTest}), and dimers
and octamers of the element at different inter-atomic distances. We see
from Table~\ref{tab:delta} that the Ultra and High basis sets have
essentially the same $\Delta$-value, indicating that they are equally
accurate. The Medium basis set is constructed by further reduction of
the High basis set, while keeping the $\Delta$-value below 4 meV.
\begin{table}
\caption{\label{tab:delta}$\Delta$-values calculated with different basis sets.
The plane-wave (PW) value is obtained using Quantum ESPRESSO with the same SG15
pseudopotentials.\cite{DeltaTest}}
\begin{ruledtabular}
\begin{tabular}{lcccc}
& Medium & High & Ultra & PW \\
\hline \cline{1-5} \\ [-2ex]
$\Delta$ (meV) & 3.45 & 1.88 & 2.03 & 1.3 \\ [-0.25ex]
\end{tabular}
\end{ruledtabular}
\end{table}

\begin{table}
\caption{\label{tab:gbrv} Summary of ATK-LCAO calculations for rock salts and
perovskites test sets. The shown RMS errors are calculated relative
to all-electron calculations. The test sets and the VASP results are taken
from Ref.~\onlinecite{Garrity2014}.}
\begin{ruledtabular}
\begin{tabular}{lcccc}
& Medium & High & Ultra & VASP \\
\hline \cline{1-5} \\ [-2ex]
Rock salt latt. const. (\%)  &  0.40 & 0.24 &  0.23 & 0.15 \\
%Rock salt bulk modulus (\%) & 11.9  & 8.28  & 10.1  & 4.5 \\
Perovskite latt. const. (\%) &  0.36 & 0.24 &  0.18 & 0.13 \\
%Perovskite bulk modulus (\%) &  6.69 & 5.67 &  4.5  & 6.1 \\
\end{tabular}
\end{ruledtabular}
\end{table}

In order to further validate the constructed SG15 pseudopotentials and
basis sets, we have performed benchmark calculations for rock salt and
perovskite crystals, as described in Ref.~\onlinecite{Garrity2014}. For each bulk
structure, the equation of state is calculated at fixed internal coordinates,
and the equilibrium lattice constant and bulk modulus are then computed.
Results are benchmarked against the scalar-relativistic all-electron calculations.\cite{Garrity2014}
Table~\ref{tab:gbrv} shows the root-mean-square (RMS) deviations from
the all-electron reference for the calculated lattice constants and bulk
moduli. For the sake of comparison, statistics for plane-wave VASP\cite{Kresse1996} calculations
is also included in this table. We see that the accuracy of the DFT calculations done
with the SG15 pseudopotentials and High (or Ultra) LCAO basis sets is
comparable to that of plane-wave calculations, while the use of the Medium basis
sets gives a slightly larger deviation from all-electron results.

We find that for typical atomistic simulations with less than 500 atoms, the LCAO calculations done with the Medium
basis set is twice as fast as that done with the High basis set,
which allows for 10 times faster LCAO calculations compared to the ones done the
Ultra basis set. Furthermore, using the Medium basis set typically permits one to do LCAO calculations an order of magnitude
faster than plane-wave calculations. In summary, the Ultra basis set
enables essentially the same accuracy of LCAO-based DFT calculations as plane-wave calculations,
at similar cost for typical 200-atom systems. Using the Medium and High basis
sets gives somewhat less accurate results of LCAO-based DFT calculations, allowing for an order of
magnitude speedup.

\subsection{Accurate semiconductor band gaps}

It is known that density functionals based on local density approximation (LDA) and generalized gradient approximation (GGA) do not allow for an accurate calculation of energy band gaps of semiconductors.\cite{MoriSanchez2008} To overcome this issue we have introduced empirical shifts of the nonlocal projectors in the SG15 pseudopotentials, in spirit of empirical pseudopotentials proposed by Zunger and
co-workers.\cite{Wang1995} The pseudopotential projector shifts (PPS) have been adjusted to reproduce technologically important properties of semiconductors such as the fundamental band gap and lattice constant. In the PPS method, the nonlocal part of the pseudopotential,
$\hat{V}_\text{nl}$, is modified in the following way
\begin{equation}
\hat{V}_\text{nl} \rightarrow \hat{V}_\text{nl} \mathrel{+}
\sum_l |p_{l} \rangle  \alpha_{l} \langle p_{l} | ,
\end{equation}
where the sum is over all projectors $p_{l}$, and $\alpha_{l}$ is an empirical
parameter that depends on orbital angular momentum quantum number, $l$. We
note that this approach does not increase the computational cost of DFT calculations.

\begin{figure}
\includegraphics[width=\linewidth]{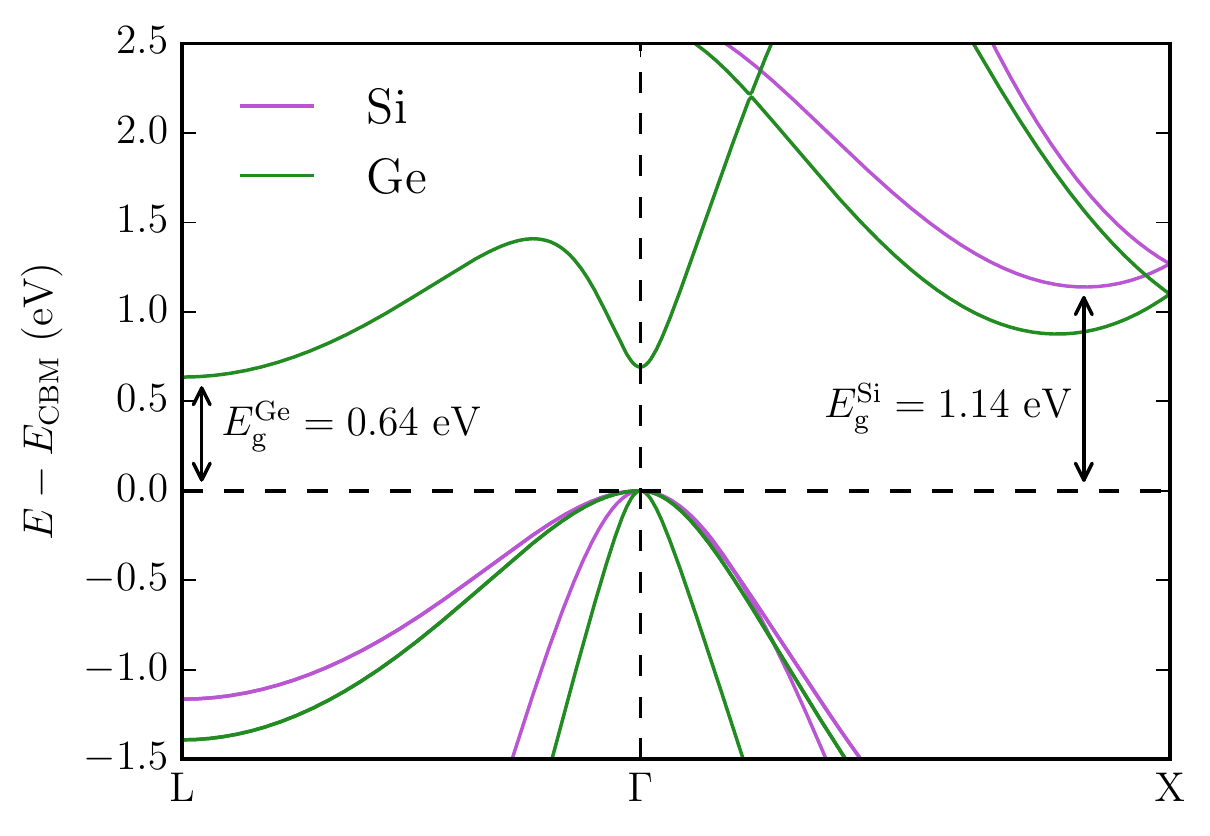}
\caption{Band structures of bulk Si and Ge, calculated with
the PPS-PBE method. Note that the fundamental $\Gamma$--$L$ band gap of bulk Ge calculated using the PBE functional is zero, in contradiction with experiment, whereas the PPS-PBE calculated band gap and other band energies agree with the experimental data in a semi-quantitative manner. The fundamental gaps for bulk Si and Ge are $E_{\mathrm{g}}^{\mathrm{Si}}=1.14$~eV and $E_{\mathrm{g}}^{\mathrm{Ge}}=0.64$~eV, respectively.}
\label{fig:bulk_bands}
\end{figure}

We have applied the PPS method in calculations for Si, Ge, and SiGe alloys, using the GGA-PBE functional\cite{Perdew1996} and the combination of SG15-pseudopotentials and Medium basis sets described in the previous section. The corresponding PPS parameters
are listed in Table~\ref{tab:pps}. Figure~\ref{fig:bulk_bands}
shows the PPS-PBE calculated band structures of bulk Si and Ge. We find
that shifting the pseudopotential projectors allows us to significant
improve not only the band gap values (compared to experiment), but also other band energies corresponding to higher conduction band valleys in Si, Ge, and related alloys.

\begin{table}
\caption{\label{tab:pps}Empirical pseudopotential projector-shifts employed
in the PPS-PBE method with the SG15-Medium combination of pseudopotential
and basis set. The shifts $\alpha_s$, $\alpha_p$, and $\alpha_d$ are applied
to $s$-, $p$-, and $d$-orbitals, respectively.}
\begin{ruledtabular}
\begin{tabular}{llll}
& $\alpha_s$ & $\alpha_p$ & $\alpha_d$ \\
\hline \cline{1-4} \\ [-2ex]
Si & $+21.33$~eV & $-1.43$~eV &           \\
Ge & $+13.79$~eV & $+0.22$~eV & $-2.03$~eV \\
\end{tabular}
\end{ruledtabular}
\end{table}

\begin{table}[b]
\caption{\label{tab:sige}}
\begin{ruledtabular}
\begin{tabular}{lcccc}
& PBE
& HSE\footnote{VASP, Ref.~\onlinecite{Schimka2011}.}
& PPS-PBE
& Exp.\footnote{From Refs.~\onlinecite{Schimka2011}, \onlinecite{Madelung2004}
and \onlinecite{Schaffler2001}.} \\
\hline \cline{1-5} \\ [-2ex]
Si \\
\hline \\ [-2ex]
a (\AA)                     & 5.468 & 5.435 & 5.443 & 5.430 \\
B (GPa)                     & 90.4  & 97.7  & 102.5  & 100.8 \\
$E_\text{g}$ (eV)           & 0.58  & 1.14  & 1.10  & 1.17  \\
$m^*_\Delta$ ($m_\text{e}$) & 0.19  &       & 0.22  & 0.19  \\ [0.5ex]
\hline \cline{1-5} \\ [-2ex]
Ge \\
\hline \\ [-2ex]
a (\AA)                     & 5.815 & 5.682 & 5.735 & 5.658 \\
B (GPa)                     & 60.9  & 71.3  & 67.4  & 77.3  \\
$E_\text{g}$ (eV)           & 0.00  & 0.72  & 0.69  & 0.74  \\
$m^*_L$ ($m_\text{e}$)      & 0.09  &       & 0.14  & 0.08  \\ [0.5ex]
\hline \cline{1-5} \\ [-2ex]
SiGe \\
\hline \\ [-2ex]
a (\AA)                     & 5.62  &       & 5.56  & 5.54 \\
B (GPa)                     & 72.0  &       & 95.3  & 86.5 \\
$E_\text{g}$ (eV)           & 0.69  &       & 0.89  & 0.97 \\
$m^*$ ($m_\text{e}$)        & 0.20  &       & 0.22  & 0.19
\end{tabular}
\end{ruledtabular}
\end{table}

Furthermore, Table~\ref{tab:sige} compares the material
parameters computed with the PPS-PBE and standard PBE approaches for Si, Ge, and a SiGe alloy to
the material parameters obtained with the computationally more expensive
HSE hybrid functional.\cite{Krukau2006} The PPS-PBE calculated lattice
constants, bulk moduli, and fundamental band gaps are in significantly
better agreement with experimentally-measured material parameters than
the parameters calculated with the PBE approach, and are
on par with HSE predictions, in general.

%
%The conduction band energies given with
%respect to the top of the valence bands are $E_{\Gamma}\! =\! 2.67$ and
%$0.69$~eV for the $\Gamma$-valley minimum, $E_{X}\! =\! 1.14$ and $0.88$~eV
%for the $X$-valley minimum, and $E_{L}\! =\! 2.96$ and $0.64$~eV for the
%$L$-valley minimum for bulk Si and Ge, respectively.

%\input{supplementary}
\end{document}